\documentclass[a4paper]{book}

\usepackage[hmarginratio=1:1,hscale=0.75,centering,bindingoffset=0mm]{geometry}
\usepackage{epsfig,wrapfig}
\usepackage{amsbsy,amssymb,amsmath,bm,ulem}
\usepackage[usenames]{color}
\usepackage{floatflt}
\usepackage{graphicx}
\usepackage{calc}
\usepackage{fancyhdr}
\usepackage{appendix}
\usepackage{listings}
\usepackage{color}

\fancypagestyle{plain}{
  \fancyhead{}
  
}

\newlength{\halfwidth}
\newlength{\tablesmall}
\newlength{\tablemedium}
\newlength{\tablelarge}
\newcommand{\f}[1]{Fig.~\ref{#1}}

\newcommand{\eqs}[2]{Eqs.~(\ref{#1}) and~(\ref{#2})}

\graphicspath{{pic/}}

\begin{document}

\title{\huge{Quantitative magneto-optical imaging with ferrite garnets}}
\vspace{2cm}
\author{\Large{Atle Jorstad Qviller}}
\maketitle

\pagenumbering{roman}

\thispagestyle{plain}
\tableofcontents
\thispagestyle{plain}
\chapter*{Abstract}

Magneto-optical imaging is a powerful technique for studying qualitative features of magnetic flux distributions in superconductors and other magnetic samples. Such distributions can have highly nontrivial features and are of interest for physicists, materials scientists and engineers. However, magneto-optical imaging does not automatically return two-dimensional maps with the actual values of the magnetic field, due to the non-linear response functions of magneto-optical indicator films and in practice also non-uniform illumination in optical setups. A quantitative treatment is needed in order to achieve such a calibration. After calibration of a magneto-optical image has been done, one can proceed to deduce the corresponding two-dimensional maps of the current density distribution from inversion of the Biot-Savart law. Some applications of quantitative magneto-optics are precise measurements of flux distributions, flux avalanches and currents in superconductors.

These notes are organized as follows: Basic aspects of magnetic flux penetration in superconductors including the Bean model are reviewed briefly in chapter \ref{beanchapter}. The rest of the notes does not strictly depend upon this chapter, however it introduces the coordinate system used and provides an useful context for some of the later discussion. In chapter \ref{experimentalchapter}, the method of magneto-optical imaging using the Faraday effect is introduced, together with practical aspects of magneto-optical experiments at low temperatures and the semi-qualitative technique of RGB addition for checking the reproducibility of flux patterns. The physics of magneto-optical indicator films and calibration of magneto-optical images into maps of magnetic field values are discussed in chapter \ref{calibrationchapter}. Next, in chapter \ref{biotsavartchapter} the Biot-Savart law and fast Fourier transformations are discussed before proceeding to inversion of magnetic field maps into current density maps. Finally, magnetometry with Faraday magneto-optical imaging is treated. Several MATLAB programs with practical implementation of the theoretical concepts discussed are available at the Github repository {\color{red}https://github.com/atlejq/Magneto-optics}.

While there has been published a large amount of work on quantitative magneto-optical imaging, the material is scattered and notation differs significantly between different authors. These notes originate from parts of the Ph. D. thesis of the author and other work done at the University of Oslo \cite{Qviller}, and collect the most important materials into a unified and updated treatment. The focus is on quantitative measurements of magnetic flux distributions in superconductors, but chapter \ref{experimentalchapter} on magneto-optical experiments at low temperatures and sections \ref{ferritesection} and \ref{calibrationsection} on the physics of indicator films and the calibration of magneto-optical raw images, respectively, have relevance outside superconductivity. For reading these notes, a knowledge of electromagnetism at the level of an undergraduate introduction course is assumed. A basic understanding of superconductivity is also useful, but not strictly required.

The author wishes to thank J\o rn Inge Vestg\aa rden for very useful comments and proofreading during the preparation of the manuscript. Additional errata found can be reported to {\color{red}atlejq@gmail.com}.

\addcontentsline{toc}{chapter}{Abstract}
\thispagestyle{plain}

\pagenumbering{arabic}
\setcounter{page}{1}

\chapter{Basic features of magnetic flux penetration in superconductors} \label{beanchapter}

\begin{figure}[h!]
 \begin{center}
  \includegraphics[width=156mm]{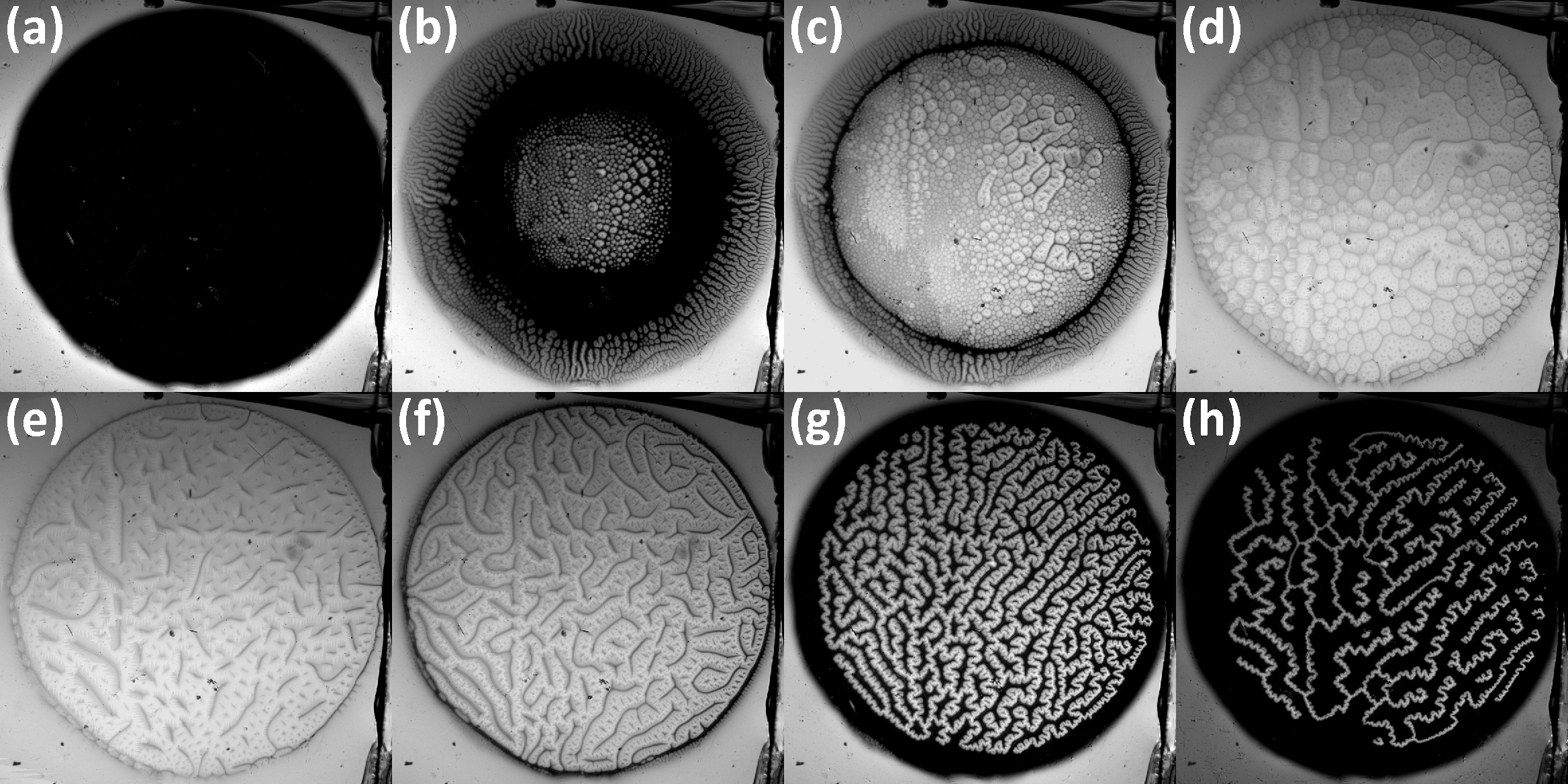}
   \caption{The intermediate state in a Pb disk of 4 mm diameter and 1 mm thickness as it is being cycled up and down in magnetic field after cooling to $T$ = 4 K in zero magnetic field. (a) $B_{a}$ = 12.9 mT ($\uparrow$), (b) 29.8 mT ($\uparrow$), (c) 34.0 mT ($\uparrow$), (d) 42.5 mT ($\uparrow$), (e) 42.5 mT ($\downarrow$), (f) 34.0 mT ($\downarrow$), (g) 17.0 mT ($\downarrow$) and (h) 8.5 mT ($\downarrow$). Up and down arrows correspond to ascending and descending fields, respectively. The pattern was gone at $B_{a}$ = 46.8 mT. }
  \label{pbdisk}
 \end{center}
\end{figure}

\noindent Superconductors are commonly categorized by their thermodynamic and magnetic properties into two types, type-I and type-II, both of which lose their electrical resistance under a \textit{critical temperature} $T_c$. In a type-I superconductor, superconductivity is destroyed above the \textit{critical field} $H_c$, but for samples with a non-zero demagnetization factor $N$, an intermediate state of normal domains containing magnetic flux mixed with superconducting domains will occur between $(1-N)H_c$ and $H_c$. An example of this is shown in the magneto-optical image series in Fig. \ref{pbdisk} of a Pb disk at $T = 4 K$, where non-trivial changes in the topological properties of the intermediate state pattern are visible as the applied field is being cycled from zero up above $H_c$ and down again to zero. In this image, light areas correspond to flux penetration, while the dark areas correspond to areas where the field is expelled by superconductivity. Type-I superconductors will not be discussed further in these notes, but it is possible to perform highly interesting quantitative investigations of their intermediate state patterns by magneto-optical imaging, see for example \cite{Prozorov}.

\begin{figure}[h!]
 \begin{center}
  \includegraphics[width=0.61 \linewidth]{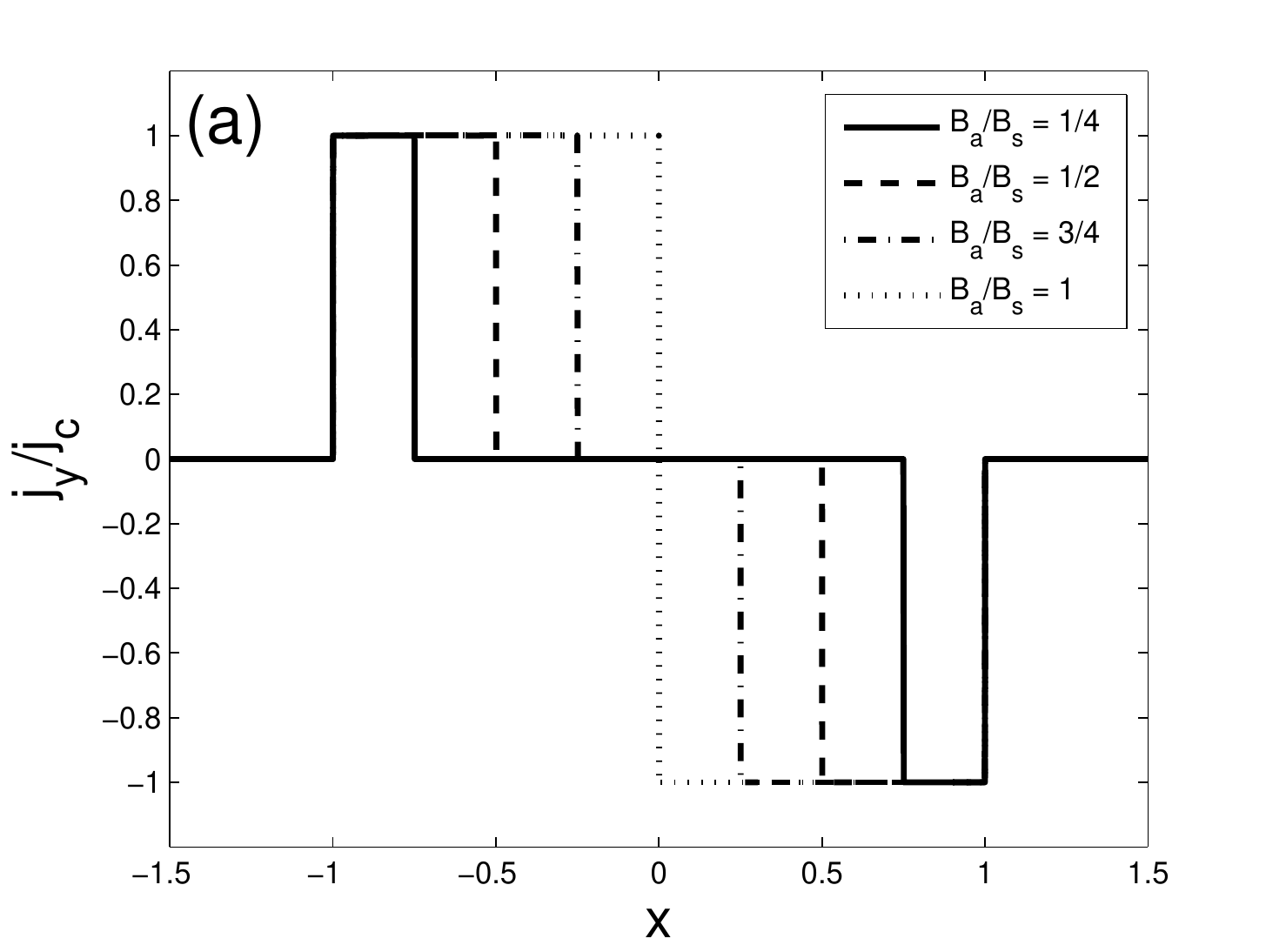}
  \includegraphics[width=0.61 \linewidth]{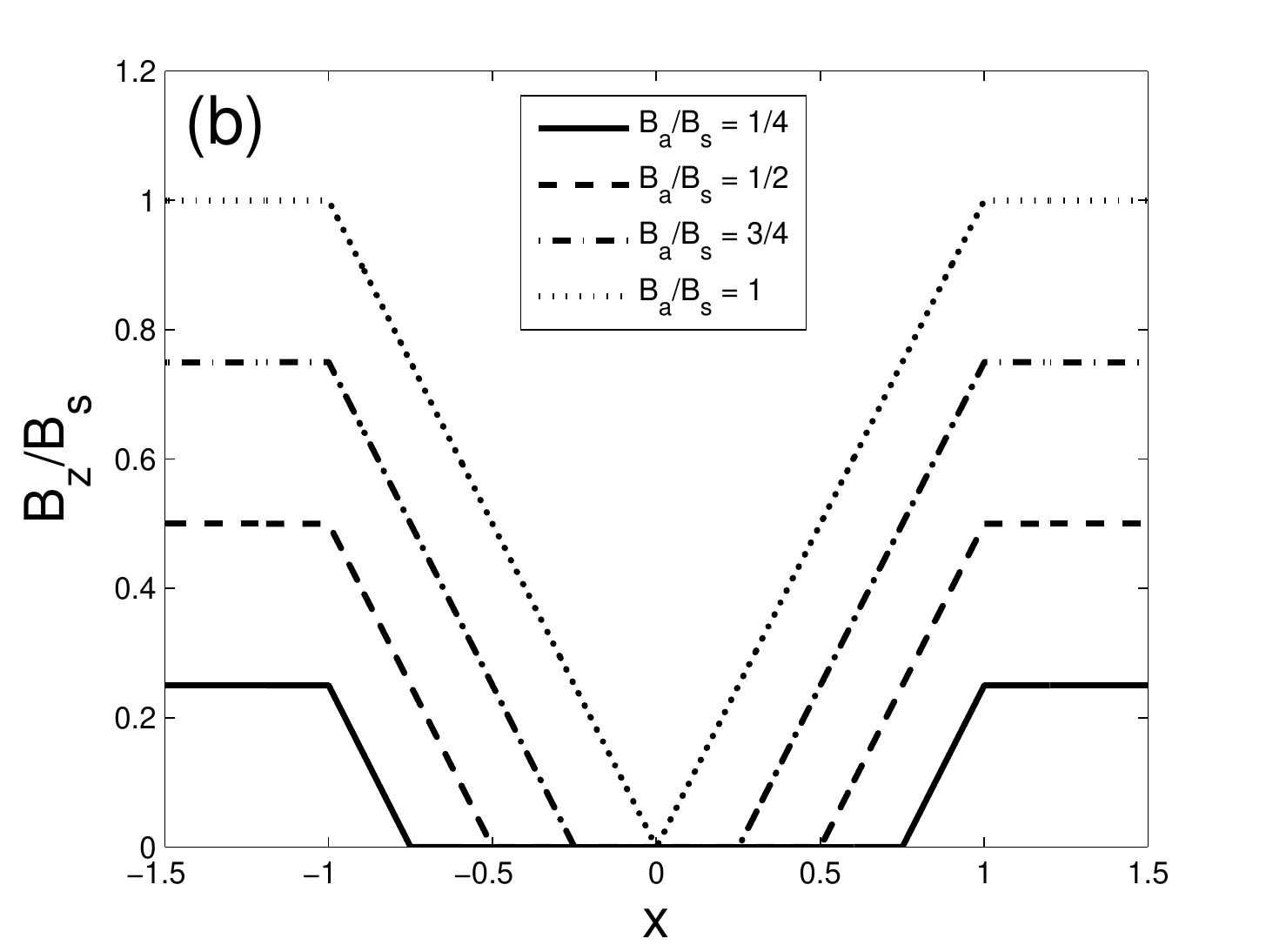}
   \caption{Solution of the Bean model in a slab geometry. The slab is centered at $x = 0$ and has width $2w = 2$. In (a) $j_y(x)/j_c$ is shown for different values of $B_a/B_s$. Corresponding $B_z(x)/B_s$ profiles are shown in (b). $\mu_0$ has been set to unity.}
  \label{slab}
 \end{center}
\end{figure}

\begin{figure}[h!]
 \begin{center}
  \includegraphics[width=0.61 \linewidth]{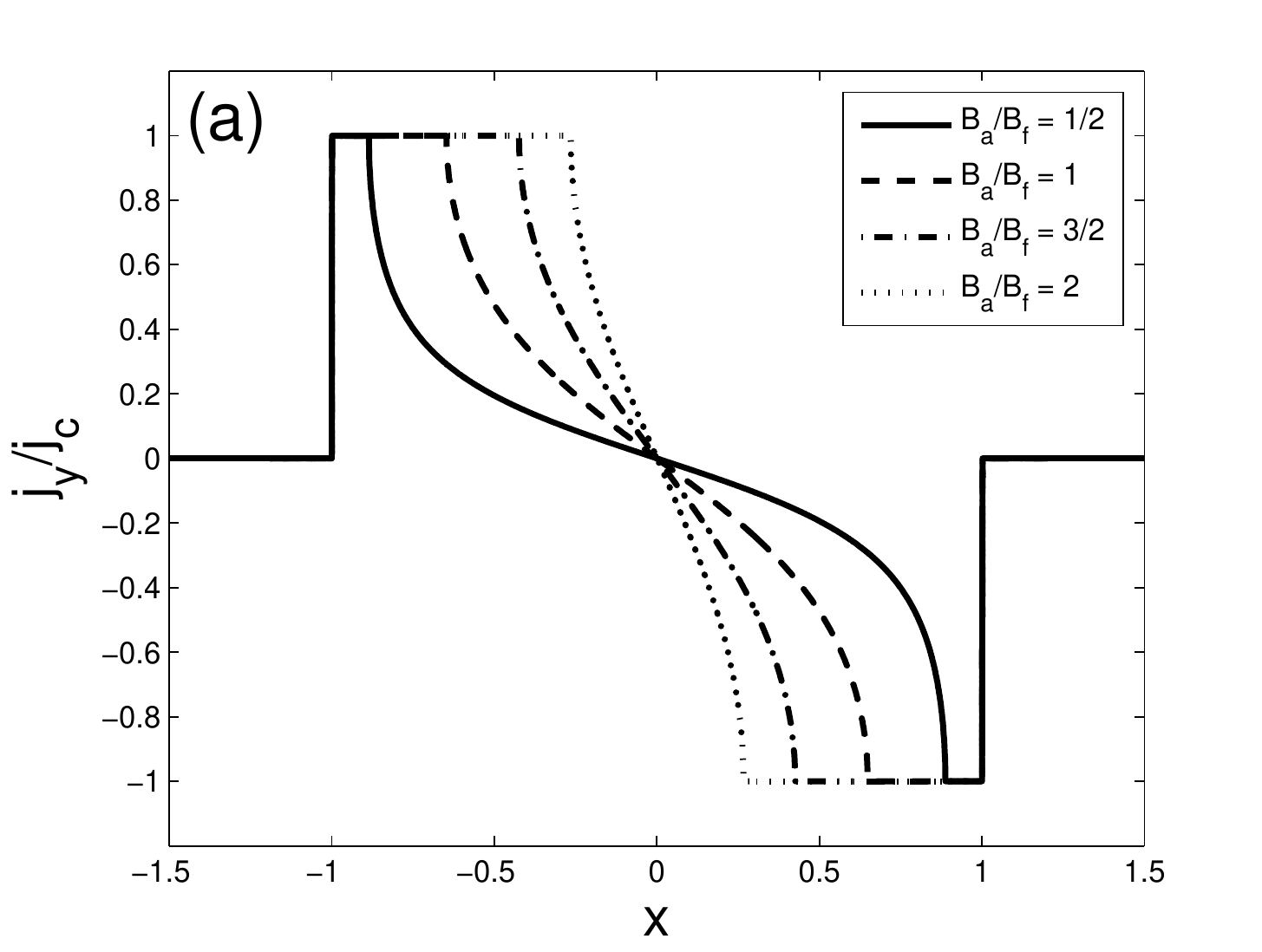}
  \includegraphics[width=0.61 \linewidth]{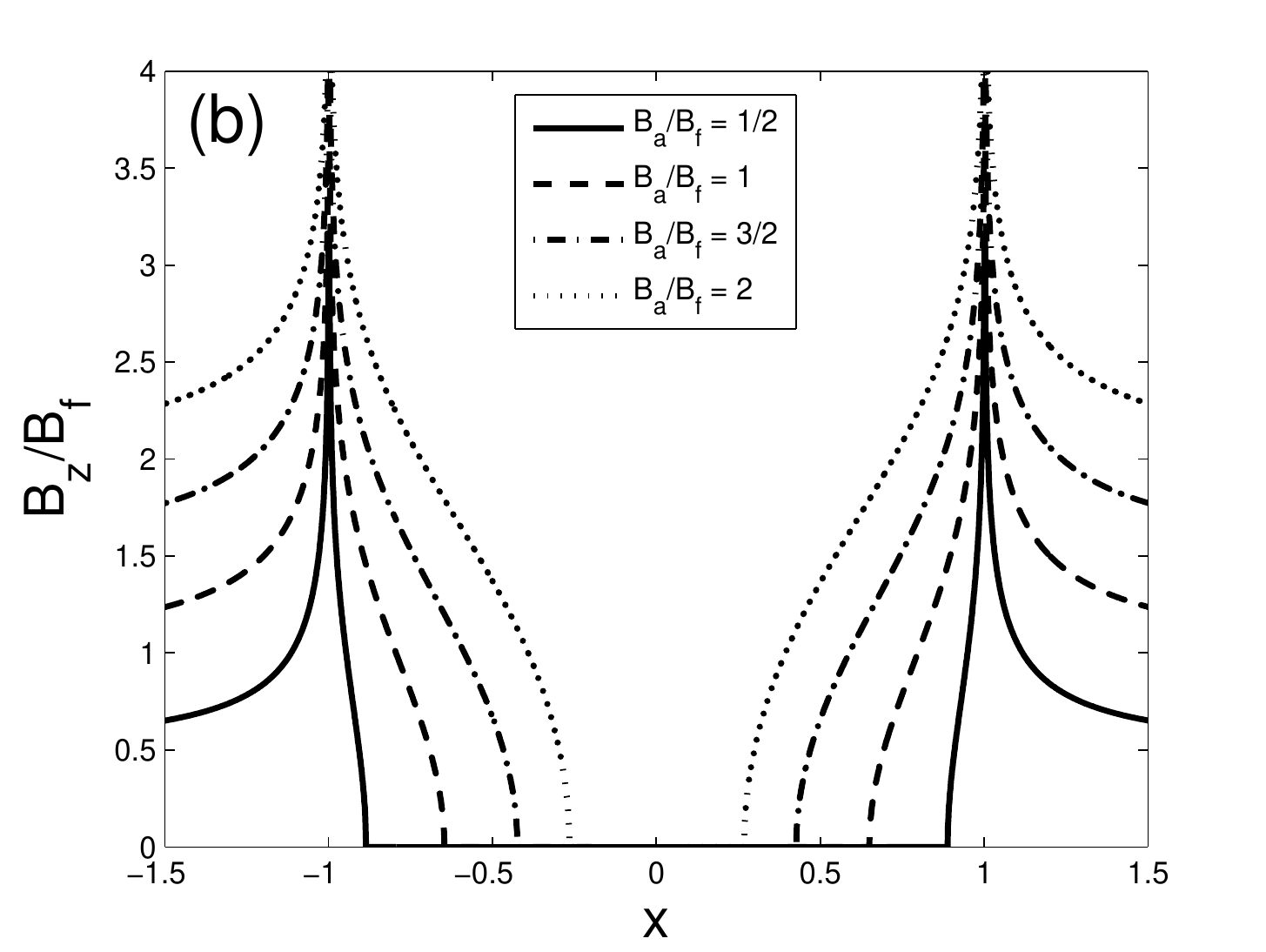}
   \caption{Solution of the Bean model in a thin film geometry. The thin film is centered at $x = 0$ and has width $w = 2$. In (a) $j_y(x)/j_c$ is shown for different values of $B_a/B_f$. Corresponding $B_z(x)/B_f$ profiles are shown in (b). The divergences in (b) at the sample edges is an artefact resulting from neglecting the first term in in Eq. \ref{ampere}. $\mu_0$ has been set to unity.}
  \label{thinfilm}
 \end{center}
\end{figure}

In a type-II superconductor, magnetic flux enter in the form of \textit{vortices} between two critical fields $H_{c_1}$ and $H_{c_2}$, each carrying one flux quantum $\phi_0$ = $h/2e$ $\approx$ 2.07 $\cdot$ $10^{-15}$ Tm$^{2}$. Apart from in a thin surface layer, superconductivity is destroyed above $H_{c_2}$, while below $H_{c_1}$ the field is expelled. Vortices in a type-II superconductor would in the absence of external forces arrange themselves into a hexagonal \textit{flux line lattice} \cite{Abrikosov}. In any physical sample, the vortices are pinned to defects like voids, grain boundaries, precipitates and dislocations, which causes deviations from the hexagonal symmetry. Vortices are affected by currents through a force given approximately as a \textit{Lorentz force density} for the case of materials with a large Ginzburg-Landau parameter $\kappa$, which includes the high-temperature superconductors (HTS) and many other materials relevant for magneto-optical imaging. The Lorentz force density is defined as
\begin{equation}
 \textbf{F}_{L} = \textbf{j} \times \textbf{B}
 \label{c1}
\end{equation}
where \textbf{j} is the current density and \textbf{B} is the magnetic flux density, both to be regarded as macroscopic, coarse-grained quantities. For low transport currents and temperatures, one can neglect thermal fluctuations and the force balance on a moving vortex is given by
\begin{equation}
 \textbf{F}_{L}+ \textbf{F}_{v}+ \textbf{F}_{p}(\textbf{B})=0
 \label{c2}
\end{equation}
Here Hall forces have also been neglected, justifiable at low temperatures, and it has been assumed that vortices are massless. $\textbf{F}_{v}$ is the \textit{viscous force density}, a parameterization of various dissipative processes affecting vortex motion and $\textbf{F}_{p}$ is the \textit{pinning force density}, generally a function of $\textbf{B}$. The pinning force on a vortex lattice is not in general equal to the sum of the pinning forces on the vortices, as the vortex lattice has internal stiffness and will have to be deformed in order to conform to the actual pinning sites. Any moving vortex or collection of vortices will induce an electric field
\begin{equation}
 \textbf{E} = \textbf{B} \times \textbf{v}
 \label{c3}
\end{equation}
In the extreme limit where the vortices are pinned so strongly that they hardly move, no viscous forces are at work and Eq. \ref{c2} is simplified to
\begin{equation}
 \textbf{F}_{L}+ \textbf{F}_{p}(\textbf{B})=0
 \label{c4}
\end{equation}
Such a situation is denoted as the \textit{critical state}. By considering Eqs. \ref{c1} and \ref{c4} and assuming a $\textbf{B}$-independent pinning force, a current corresponding to the maximum pinning force density can be defined as
\begin{equation}
 F_{p} = j_{c}B
 \label{c5}
\end{equation}
The internal current that characterizes this situation is called the \textit{critical current} $j_c$. In a model by Bean \cite{Bean62}, a field-independent critical current is postulated, while extensions by Kim \cite{Kim63} and others more realistically allows the critical current to be field dependent, $j_c(B)$. In the original \textit{Bean model} it is assumed that a constant current equal to $j_c$ is flowing in areas of the superconductor where magnetic flux has penetrated and has formed a flux density gradient. In addition, zero current density in flux-free regions of the sample, or when magnetic flux has been frozen in during a \textit{field cooled} experiment, is assumed. When there is no vortex motion, there is also no dissipation of energy, and any current below $j_c$ can theoretically flow without any resistance. Reversible properties of the magnetization are ignored in the simplest version of the Bean model and if again a large $\kappa$ is assumed, one may set $B = \mu_0 H$ with little error.
\begin{figure}[h!]
 \begin{center}
  \includegraphics[width=0.6 \linewidth]{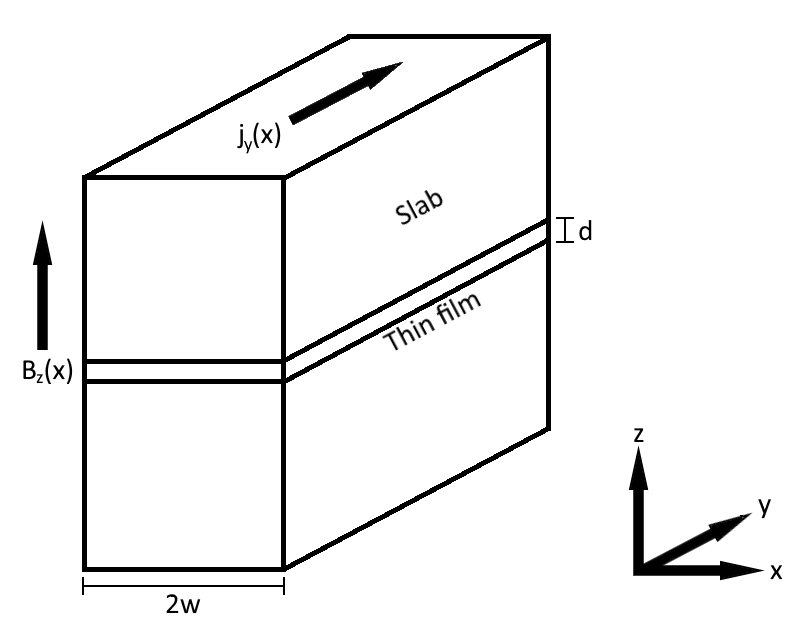}
   \caption{Slab and thin film geometries. Both the slab and the thin film are infinite in the $\pm y$-directions, and the slab is additionally infinite in the $\pm z$-directions. The centre of the slab or film corresponds to the origin of the coordinate system.}
  \label{slabfig}
 \end{center}
\end{figure}
\noindent The expression for the magnetic field in the $z$-direction $B_z$ at depth $x$ from the edge inside a superconducting slab or thin film in the Bean model can be derived from Amp\`{e}re's law. Both the slab and the thin film are furthermore assumed to have infinite extent in the $y$-direction and a width $2w$ in the $x$-direction, see Fig. \ref{slabfig} for a sketch of the geometry. In the following, it is additionally assumed that $\lambda << 2w$, where $\lambda$ is the London penetration depth of the superconductor.  When the magnetic field is applied only along the $z$-direction and currents only along the $y$-direction, Amp\`{e}re's law reads
\begin{equation}
 \frac{\partial B_z}{\partial x} - \frac{\partial B_x}{\partial z} =-\mu_0 j_y(x)
 \label{ampere}
\end{equation}
For the slab geometry, infinite extent in the $z$-direction is also assumed. $B_x$ vanishes then by symmetry and Eq. \ref{ampere} is simplified to
\begin{equation}
 \frac{\partial B_z(x)}{\partial x} =-\mu_0 j_y(x)
 \label{ampereslab}
\end{equation}
Starting from the virgin state with $B = 0$ everywhere, Eq. \ref{ampereslab} is trivial to integrate with $j_y$ = $j_c$ and $B_z = B_{a}$ as boundary condition, where $B_a$ is the applied field
\begin{equation}
 B_z(x) = B_{a} -\mu_0 j_{c} x
 \label{c7}
\end{equation}
\noindent Thus, the magnetic flux density falls off with a constant gradient inside the sample, with slope $\mu_0 j_{c}$. For $B_s = \mu_0 j_{c}w$ the slab is fully penetrated, defining a characteristic applied field for a the slab geometry, and the current density has reached $j_c$ in the entire sample. More formally, the current densities and magnetic field profiles inside a slab with edges at $x = \pm w$ can be written as
\begin{equation}
    j_y(x) = \left\{ \begin{array}{ll}
      j_{c} &\mbox{, $-w < x < -a $} \\
       0  &\mbox{, $|x| < a $} \\
       -j_{c} &\mbox{, $a < x < w  $}
       \end{array} \right.
 \label{3currslab}
\end{equation}
\begin{equation}
    B_z(x) = \left\{ \begin{array}{ll}
      0 &\mbox{, $0 \leq |x| < a $} \\
       \mu_0 (|x|-a) j_{c}  &\mbox{, $a \leq |x| < w $} \\
       B_a &\mbox{, $w \leq |x|$}
       \end{array} \right.
 \label{3fieldslab}
\end{equation}
\noindent where a flux-free region of width $2a$ is present in the center of the sample with $a = w(1-B_a/B_s)$. In Fig. $\ref{slab}$ (a) and (b), plots of Eqs. $\ref{3currslab}$ and $\ref{3fieldslab}$, respectively, for different values of $B_a/B_s$ are shown. If $B_a$ is ramped up from zero and then down to zero again, flux will be trapped inside the sample, resulting in hysteresis and a double pyramid-shaped $B_z(x)$ for the slab geometry. This state is called the \textit{remanent state}.
\noindent In a thin film of the same width $2w$ as the slab, infinite length in the $\pm y$-directions, but now thickness $d$ in the $z$-direction, as sketched in Fig. \ref{thinfilm}, the second term in Eq. \ref{ampere} dominates instead of the first and nonlocal electrodynamics give a different solution than for the slab geometry (in the following $\lambda < d << w$ is assumed) \cite{Norris, Brandt93, Zeldov}. The magnetic field profiles are still characterized by a single parameter $j_{c}$ in addition to geometric factors, but do not have constant slope anymore. Thus, by applying a homogenous magnetic field to a superconducting thin film, an inhomogenous field is generated above it. The solution for the current densities (to be considered as averages across the film thickness $d$) and magnetic field profiles in the strip plane are now given by

\begin{equation}
    j_y(x) = \left\{ \begin{array}{ll}
      j_{c} &\mbox{, $-w < x \leq -a $} \\
       -\frac{2j_{c}}{\pi} arctan \Big(\frac{x}{w} \sqrt{\frac{w^{2} - a^{2}}{a^{2}-x^{2}}}\Big)  &\mbox{, $-a < x < a $} \\
       -j_{c} &\mbox{, $a \leq  x < w  $}
       \end{array} \right.
 \label{3currfilm}
\end{equation}
\begin{equation}
  B_z(x) = \left\{\begin{array}{ll}
      0 &\mbox{, $-a \leq x \leq a $} \\
       \frac{\mu_{0} j_{c} d}{\pi} ln \frac{|x| \sqrt{w^{2}-a^{2}} + w \sqrt{x^{2}-a^{2}}} {a \sqrt{|x^{2}-w^{2}|}} &\mbox{, $a < |x| $} \\
       \end{array} \right.
 \label{3fieldsfilm}
\end{equation}
where $2a$ is again the width of the field-free region in the middle of the thin film, but now with
\begin{equation}
 a = \frac{w}{cosh(\frac{B_{a} \pi }{\mu_{0} j_{c} d})} = \frac{w}{cosh(\frac{B_{a}}{B_f})}
 \label{c9}
\end{equation}
In contrast to the slab solution, $a$ is now a non-linear function of $B_a$. The quantity $B_f = \mu_{0}j_{c}d/\pi$ defines a characteristic field for the thin film geometry, and in Fig. $\ref{thinfilm}$ (a) and (b), plots of Eqs. \ref{3currfilm} and \ref{3fieldsfilm} are shown, respectively, for different values of $B_a/B_f$. Eq. \ref{c9} can be used to estimate the critical current directly from a magneto-optical image of a sample on the assumption that the current is approximately field independent if $d$ is known. An example of such an estimation is shown in section \ref{magnetic} on magnetic field to current inversion. Note that there is no current-free region in thin films exposed to an external field as there is in the corresponding current distribution for a slab, this is seen by comparing Eq. \ref{3currfilm} to Eq. \ref{3currslab}. For $|x|>a$, $|j_y| = j_c$ in both geometries.

No pinned vortex lattice is completely static at temperatures different from the absolute zero, and including thermal activation will cause \textit{flux creep}. Thermal effects also cannot be neglected unless $j<<j_c$. Additionally, the analysis above is not applicable for non-static situations and a slow, quasi-static application of $B_a$ is assumed. Alternating currents will not flow without resistance even if their magnitude is less than $j_c$ due to viscous forces. In practice, there is also some electrical resistance below $j_c$. The relation between electric field and current in the superconductor is often parameterized in the form
\begin{equation}
  \textbf{E}=\rho_{0} (\frac{j}{j_{c}})^{n-1}\textbf{j}
 \label{c10}
\end{equation}
\noindent This power law E-j relation is one of several \textit{material laws} of the superconductor. It reduces to Ohm's law if $n=1$ and to the Bean model if $n=\infty$. The Bean- and other critical state models can also be extended to handle samples with anisotropic $j_c$. In this case, different currents will flow in the $x$- and $y$-directions of the sample. Current conservation inside the sample is expressed as the current having zero divergence
\begin{equation}
  \nabla \cdot \textbf{j}= 0
 \label{c11}
\end{equation}
and for a rectangular sample, this requires that the current turns around at an angle
\begin{equation}
  \alpha = \arctan \frac{j_{c,y}}{j_{c,x}}
 \label{c12}
\end{equation}
\noindent $j_{c,x}$ and $j_{c,y}$ are the critical currents in the $x$- and $y$-directions, respectively. \f{ybco8} shows a magneto-optical image of a fullypenetrated superconducting thin film with anisotropic $j_c$. The light areas of this image inside the sample area again correspond to penetrated magnetic flux, while the dark lines show up where currents change abruptly direction and the applied field therefore is expelled. They are known as \textit{discontinuity lines} or simply \textit{d-lines}\footnote{To be a little more pedantic, there are actually two different kinds of d-lines, both visible in \f{ybco8}. The dark, internal lines where the current $\textbf{j}$ changes direction, but $|j|$ is constant, are denoted d$^{+}$ lines. In contrast, the bright lines at the three visible edges of the sample are denoted d$^{-}$ lines. d$^{-}$ lines appear where $|j|$ abruptly changes \cite{Schuster}.}. In a flat sample with isotropic currents in the $x$- and $y$-directions, the d-lines are situated at an angle of $45^{\circ}$ from the sample edge. In a sample with anisotropic in-plane currents, such as the one shown in \f{ybco8}, this is no longer the case. Here, the angle $\alpha$ of the d-lines relative to the sample edge along the $x$-axis is approximately $60^{\circ}$, giving a ratio of currents $j_{c,y}/j_{c,x} \approx 1.7$ by Eq. \ref{c12}. The angle $\alpha$ is generally magnetic field- and temperature dependent, as the magnitudes of the critical currents in different directions may have different magnetic field- and temperature dependencies.
\begin{figure}[t!]
 \begin{center}
  \begin{tabular}{lll}
  \includegraphics[width=0.6 \linewidth]{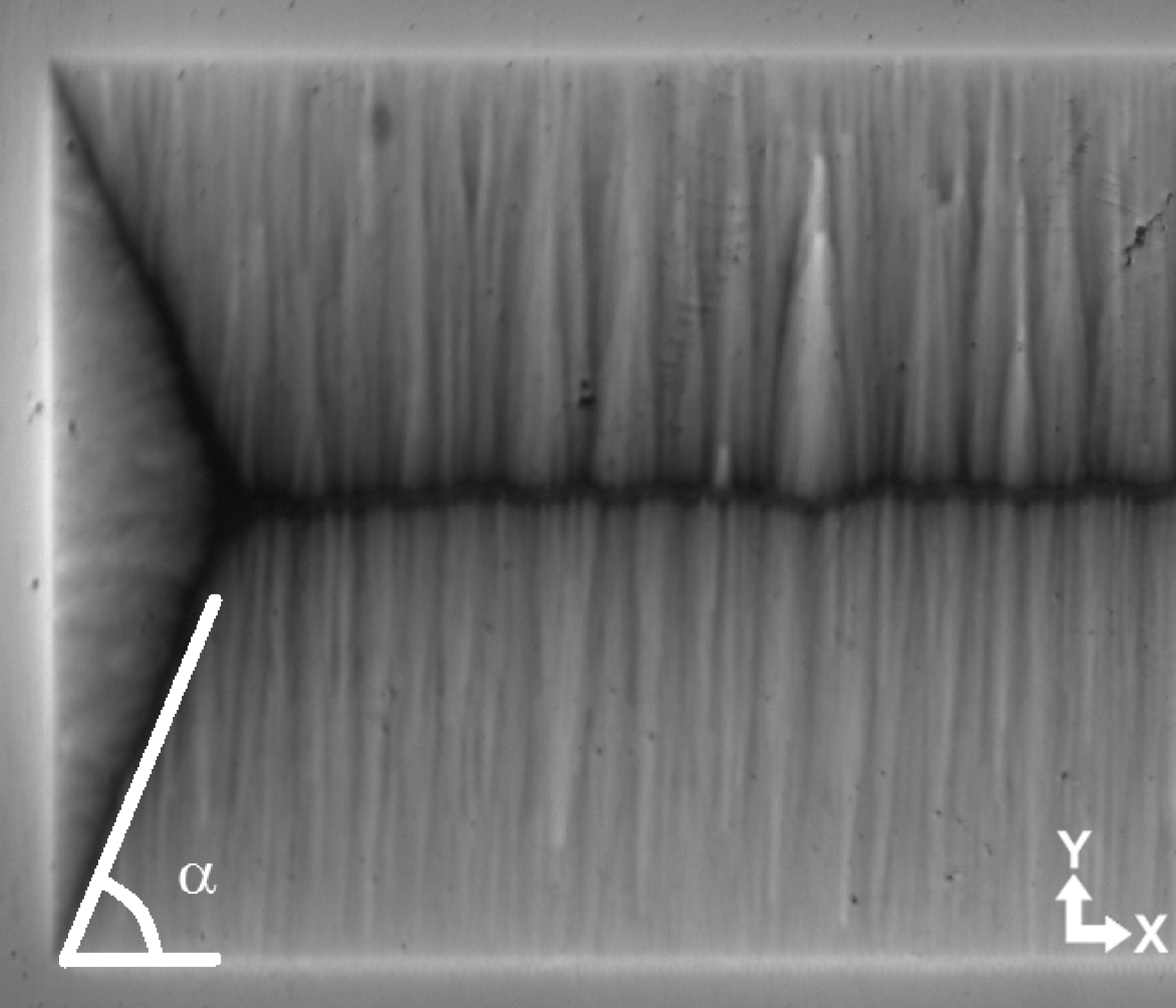}
  \end{tabular}
   \caption{A magneto-optical image of a YBCO strip grown on a substrate with tilt angle of $8^{\circ}$, captured at $T = 6$ K and $B_a = 22.5$ mT after cooling in zero magnetic field. The d-line angle $\alpha$ relative to the sample edges deviates from $45^\circ$ and thus indicates anisotropic $j_c$. The strip is approximately 1 mm wide.}
  \label{ybco8}
 \end{center}
\end{figure}
MATLAB code for plotting profiles of magnetic field and current distributions for the Bean model in slabs and thin films is given in {\color{red}beanmodel.m}.

\chapter{Magneto-optical experiments} \label{experimentalchapter}

In this chapter, a brief description of the experimental method of magneto-optical imaging using the Faraday effect is given. Afterwards, practical considerations when doing low-temperature magneto-optical experiments using cryostats are discussed.

\section{Magneto-optical imaging using the Faraday effect}
Several techniques can be used to visualize magnetic field distributions. Magneto-optical imaging based upon the Faraday effect has several advantages such as locality, extremely good time resolution (sub-nanosecond) and good spatial resolution (typically a few $\mu$m). In addition, the technique is non-destructive to the sample in contrast to methods such as Bitter decoration.

\begin{figure}[h]
 \begin{center}
   \includegraphics[width=0.5 \linewidth]{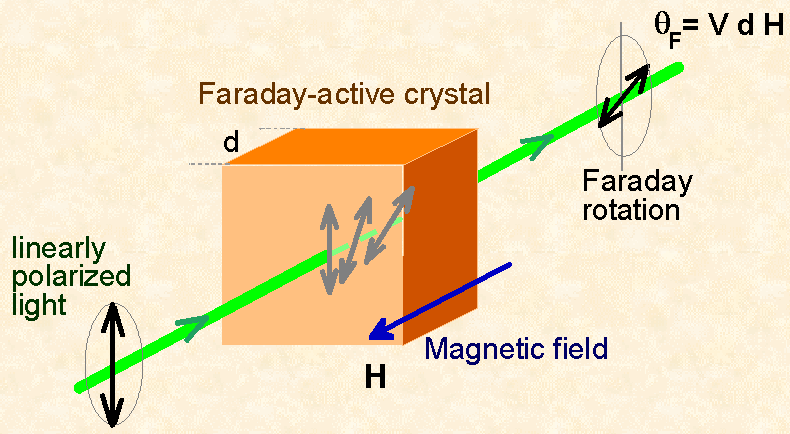}
   \caption{Rotation of the polarization vector of light in a Faraday active crystal.}
  \label{faradayeffect}
 \end{center}
\end{figure}

\noindent The Faraday effect is the rotation of the polarization vector of light in a transparent medium when a magnetic field is applied along the wave vector of the light, as seen in Fig. \ref{faradayeffect}. For paramagnetic materials or at the virgin curve for spontaneously magnetized materials, the rotation can at small fields be written as the product of a material parameter called the Verdet constant $V(\omega)$, which is generally a function of the light wavelength $\omega$, the magnetic field component parallel to the light beam and the length of the optical path inside the material.
\begin{figure}[h]
 \begin{center}
   \includegraphics[width=0.5 \linewidth]{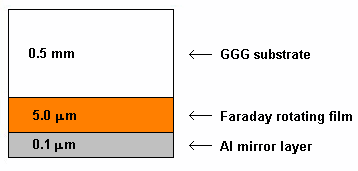}
   \caption{Structure of a Faraday-rotating indicator film with three layers, not to scale.}
  \label{film}
 \end{center}
\end{figure}

Many materials including superconductors do not show a significant magneto-optical Faraday (or Kerr effect) by themselves and this necessitates the use of a magneto-optical indicator film placed directly on top of them if magneto-optical imaging is to be applied. Several types of indicator films have been used for magneto-optical Faraday imaging including europium selenide (EuSe), ferrite garnets with out-of-plane magnetization and ferrite garnets with in-plane magnetization. EuSe films have to be deposited on the sample and have a highly temperature dependent Faraday rotation. Ferrite garnets with out-of-plane magnetization show an awkward bubble or labyrinth domain structure which limits their applicability. This work utilizes ferrite garnets with in-plane magnetization for the Faraday rotating indicator film, which, depending on exact chemical composition, can be used up to at least $B_a = 70-100$ mT. They generally have the highest Faraday rotation for green light. The indicator film is made of three layers: An aluminium mirror, a Faraday-rotating ferrite garnet layer and a gadolinium gallium garnet (GGG) substrate of a specific orientation, see Fig. \ref{film} for a sketch. These GGG substrates are not regularly produced anymore and are starting to be hard to come by.
 \begin{figure}[h]
 \begin{center}
   \includegraphics[width=0.43 \linewidth]{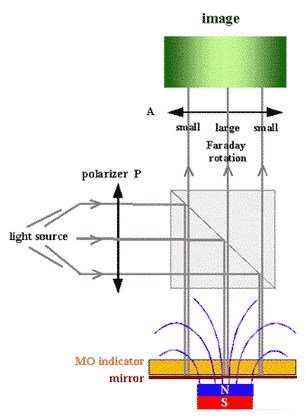}
   \caption{Schematic setup of a magneto-optical polarization microscope. The sample is symbolized by a magnet below the indicator film.}
  \label{faradaymic}
 \end{center}
\end{figure}

It is possible to construct a polarization microscope utilizing the Faraday effect, as sketched in Fig. \ref{faradaymic}. The beam from the light source is first linearly polarized by a polarizer (P). It is then reflected by a beam splitter with a partially transparent mirror towards the Faraday rotating indicator film. The polarization of the light beam then undergoes a Faraday rotation proportional to the local magnetic field, hits a mirror deposited on the indicator film directly above the superconducting or magnetic sample and is reflected, undergoing an equivalent Faraday rotation in the same direction as the first and is thereafter transmitted back though the beam splitter. In order to generate the image, the light is finally transmitted through a second polarizer (A), often denoted the \textit{analyzer} and is then seen in the eyepieces of the microscope or captured by a CCD camera.

If the angle between the polarizer and the analyzer is $90^{\circ}$, a configuration called \textit{crossed polarizers}, no light that has not undergone a Faraday rotation in the indicator film will in theory get through the analyzer. Dark areas in the image will then correspond to no Faraday rotation and thus no magnetic field parallel to the optical path toward the analyzer and the camera, while brighter areas correspond to Faraday rotation due to magnetic field in the area. However, as will be discussed in the chapter about quantitative magneto-optics, this configuration has the lowest sensitivity. For higher sensitivity one can instead operate the setup at a nonzero angle between the polarizer and the analyzer.

The microscope will usually include a cryostat and a magnetic solenoid around the sample, so that samples can be investigated at low temperatures and various magnetic fields. A comprehensive review of magneto-optical imaging utilizing the Faraday effect is given in ref. \cite{Jooss}.

\section{Practical magneto-optical experiments at low temperatures}
In order to investigate superconducting or magnetic samples at low temperatures, one has to mount and cool them in a cryostat. Two types of cryostats are discussed in this text: Helium flow cryostats, where the cooling is implemented through the evaporation of a helium flow and closed-cycle cryostats that are essentially a refrigerator using helium. The latter are becoming popular due to helium shortages.

\noindent Before mounting the sample in a cryostat, samples and indicator films should be inspected under an ordinary stereomicroscope with a few times magnification. One will then easily see dust, grease or other kinds of dirt. If the sample is dirty, it should be cleaned in toluene and wiped clean with soft, lint-free wipes. It is very important to note that even small scratches on superconducting samples often render them useless, as the flux distributions below $T_{c}$ can be severely disturbed. It is also easy to damage the mirror deposited on indicator films. Samples should be kept in an exicator with silica gel to absorb moisture to avoid degradation and oxidation. One can additionally evacuate the exicator for air to further reduce the chance of degradation, the only exception is maybe perovskite HTS materials like YBCO, where the oxygen stoichiometry can possibly be disturbed by outgassing into the vacuum.

\begin{figure}[t]
 \begin{center}
   \includegraphics[width=0.5 \linewidth]{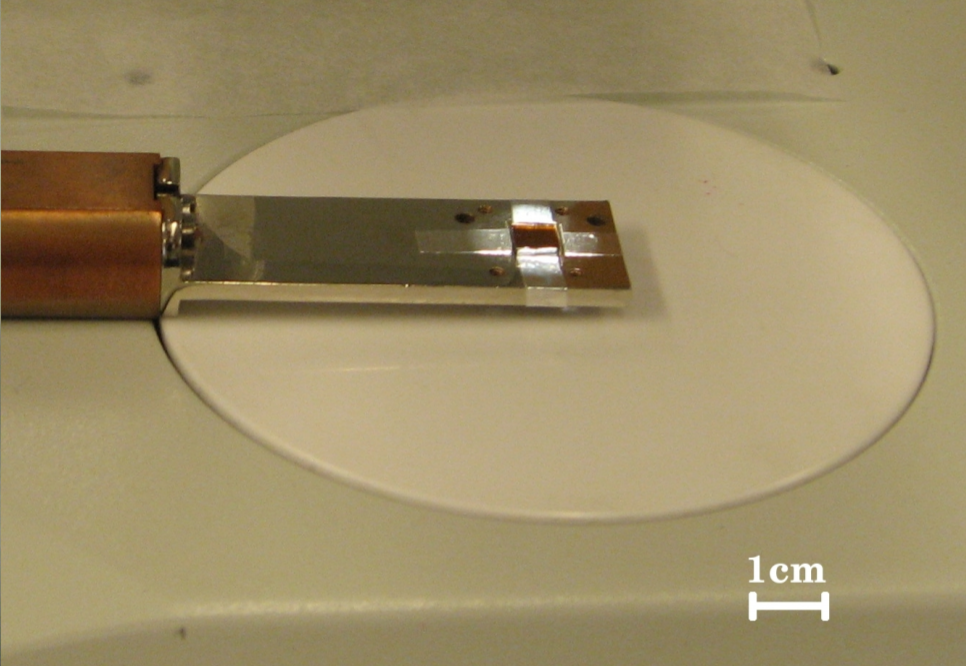}
   \caption{Cold finger, magneto-optical indicator film and aluminum L-shapes. A NbN sample is mounted below the indicator. Figure from ref. \cite{Eliassen}}
  \label{coldfinger}
 \end{center}
\end{figure}

Samples are mounted on a cold finger. Vacuum grease will fix the sample physically and ensure a good thermal contact between the cold finger and the sample. Too much grease will act as a thermal insulator. The magneto-optical indicator film is placed directly on top of the sample, and it is important that it is not tilted relative to it, as partial blurring of the image will result. Blurring will also occur if the sample is not flat. The sample must be restrained to not slip away from the sample as a result of vibrations. 4 L-shaped pieces of aluminum tape will ensure this, as shown in Fig. \ref{coldfinger}.

A radiation shield of aluminum is attached around the cold finger before it is placed in the cryostat. This shield and all other metal surfaces should be free of fingerprints and grease, as these are opaque and radiate in the infrared range. It is very important to avoid getting vacuum grease between the indicator film and the sample. At low temperatures, the vacuum grease will solidify and stress the indicator film. This causes ”stripe domains” to appear and ruins the image. Also, the grease should be evenly distributed below the sample. If there is air trapped below the sample, sudden displacements can result from thermal volume changes. Different kinds of vacuum grease exist, for example Dow Corning, Apiezone and Cryocon grease from ARS. The first is fairly standard and works under most circumstances. Apiezone has better thermal conductivity, but is harder to distribute evenly as it is more sticky. The Cryocon grease is maybe the best, it contains tiny copper particles to ensure an even better thermal conductivity. In a closed-cycle cryostat, this is important, as its cooling power is much smaller than the cooling power of a flow cryostat.

One can generally get very good images with 20x magnification. 50x magnification is attainable, but these images may be dark and suffer from depolarization effects. The higher the magnification, the worse are vibration effects and vibration insulation becomes more and more important. Even a person walking in a lab will disturb the image at 20x. For closed-cycle cryostats with exchange gas, the vibration insulation is very good and the small vibrations seen are mostly coming from the vacuum pump. Flow cryostats are somewhat more plagued by vibrations coming from both the vacuum pump and the helium pump.

In order to transfer helium gas from a Dewar tank to a flow cryostat, a transfer tube is used. It must be slowly inserted into the Dewar tank to let the helium replace the air inside (obviously the valve on top of the transfer tube must be open). Otherwise a blockage of frozen air will result. If this happens either during insertion or the experiment, the tube must be taken out and dried, preferably with a heat gun. One must also run air through it to dry up any moisture inside of it. Flow cryostats should not be cooled faster than 10 $^{\circ} \mathrm{C}$ per minute, as excessive thermal stress can degrade the materials in the setup in the long run. Such concerns are minimal for closed-cycle cryostats, as they are not able to cool nearly that fast. When heating any cryostat up again, air must not be allowed to enter the vacuum inside it before the temperature has reached room temperature. Otherwise, internal condensation will result, possibly causing degradation of the setup. Air cooled magnet coils must not be allowed to overheat. As a rule of thumb, they should not become hotter than what is comfortable to touch (50 $^{\circ} \mathrm{C}$). This is a general problem for any experiment using a high magnetic field for extended periods of time. Water cooling may be implemented, but will make the entire setup more bulky and impractical.

Light sources in the form of lamps should be switched on and off as little as possible, as this shortens their lifespan. Lamps running of the AC mains will induce a spurious flickering in the images. Although this effect is only about a few percent, it would be preferable to use DC powered green high-intensity LEDs. A green laser can also be used, but will require more safety precautions and introduce interference noise. Optical components must be carefully selected to not induce large depolarization effects, and those in contact with magnetic fields such as objective lenses must also have small Verdet constants.

\section{RGB addition of images}
\begin{figure}[h!]
 \begin{center}
  \includegraphics[width=0.495 \linewidth]{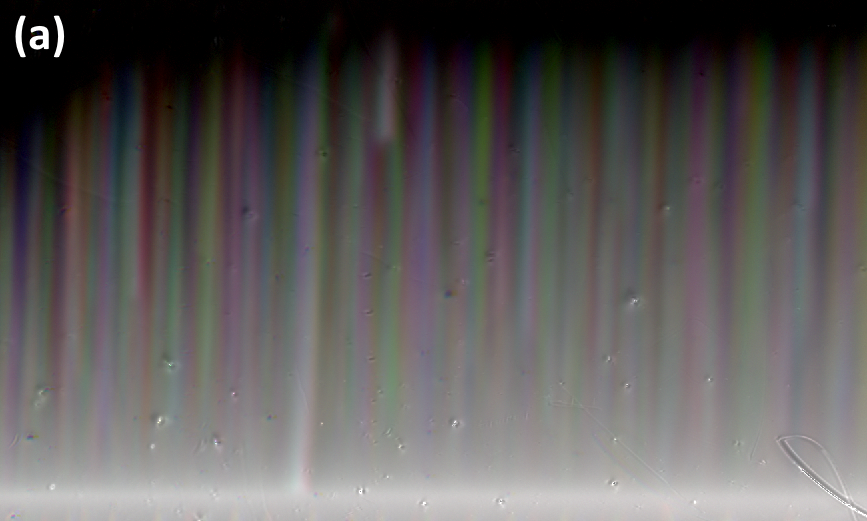}
  \includegraphics[width=0.495 \linewidth]{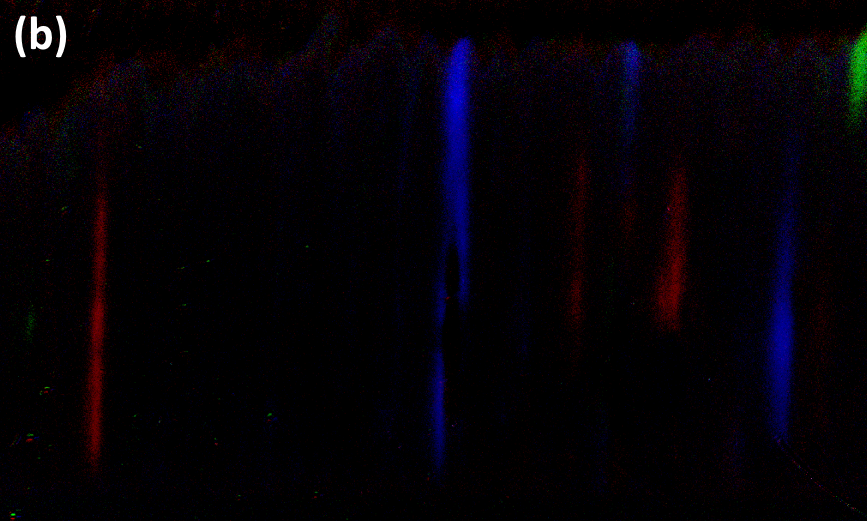}
  \caption{(a) RGB addition of three magneto-optical images obtained at $B_{a}$ = 17 mT and $T$ = 4 K of a YBCO film on a $14^{\circ}$ tilted substrate. A needle-like flux pattern is seen to be mostly, but not completely repeatable. (b) RGB addition of differential images corresponding to (a) where the images captured at $\Delta B = 42.5$ $\mu$T lower field has been subtracted. Several quasi-one-dimensional flux avalanches are visible. These are of a stochastic nature and therefore do not add up to the color white.}
  \label{rgb}
 \end{center}
\end{figure}

From additive color mixing we know that the sum of the colors red (R), green (G) and blue (B) is white. This enables us to check the reproducibility of flux patterns and flux avalanches by adding up three series of either ordinary or differential magneto-optical images, respectively, obtained under identical experimental conditions (same cooling procedure, $T$ and $B_a$). Differential images are made by capturing an image, ramping up the applied magnetic field $B_a$ by an amount $\Delta B$, capturing a second image and then subtracting the first image from the second in order to observe changes in the flux pattern. Examples of RGB addition are shown for ordinary images in Fig. \ref{rgb} (a) and for differential images in Fig. \ref{rgb} (b). Note that when working in MATLAB, offsets may have to be introduced to avoid clipping of negative flux changes in differential images, as images are represented by positive numbers. RGB addition and taking differential images do not require calibration of the magneto-optical images investigated and can be considered to be a semi-qualitative technique.

MATLAB codes for RGB addition of ordinary and differential images are included in {\color{red}RGBord.m} and {\color{red}RGBdiff.m}, respectively.

\chapter{Calibration of magneto-optical images} \label{calibrationchapter}
Qualitative magneto-optical imaging of superconductors can locally and in real-time reveal features of the flux penetration pattern, like penetration depths, defects, anisotropic penetration and also dynamic features like flux creep and flux avalanches. However, for actually measuring magnetic field values with magneto-optical imaging, a quantitative treatment is required, which is a much more complicated task. It involves calibrating the measured light intensity from the indicator as a function of local magnetic field. This chapter starts with a discussion of the physics of magneto-optical indicator films and proceeds to calibration of magneto-optical images.

\section{Ferrite garnets with in-plane magnetization} \label{ferritesection}
In this work the focus is on magneto-optical indicator films made of ferrite garnets with in-plane magnetization, which in use shows a characteristic and sometimes problematic sawtooth-pattern of magnetic domains. Their advantages include a high and rather constant Faraday rotation in a temperature range from the lowest attainable temperatures in a Helium-4 cryostat to above the $T_{c}$ of HTS materials. The spontaneous in-plane fields $B_{A}$ of these garnets is typically at least 100 mT, but if a too large field is applied, the rotation saturates. Unfortunately, the in-plane magnetized garnets are also sensitive to the in-plane field components $B_{x}$ and $B_{y}$, which will cause a non-locally reduced Faraday rotation and thus underestimation of the local $B_{z}$ \cite{Johansen}.

\begin{figure}[h!]
 \begin{center}
  \begin{tabular}{lll}
   \includegraphics[width=0.8 \linewidth]{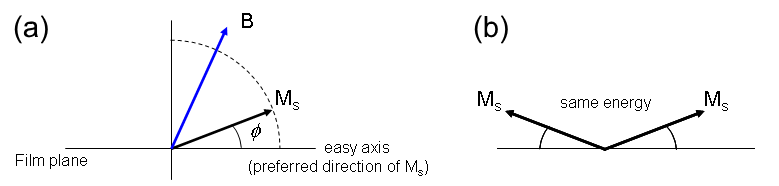}
   \end{tabular}
   \caption{(a) Coordinate system showing orientations of $\textbf{B}$, $\textbf{{M}}_{s}$ and $\phi$. (b) Symmetry of the interaction energy in an external field.}
  \label{indicatorfig1}
 \end{center}
\end{figure}

\begin{figure}[h]
 \begin{center}
   \includegraphics[width=0.5 \linewidth]{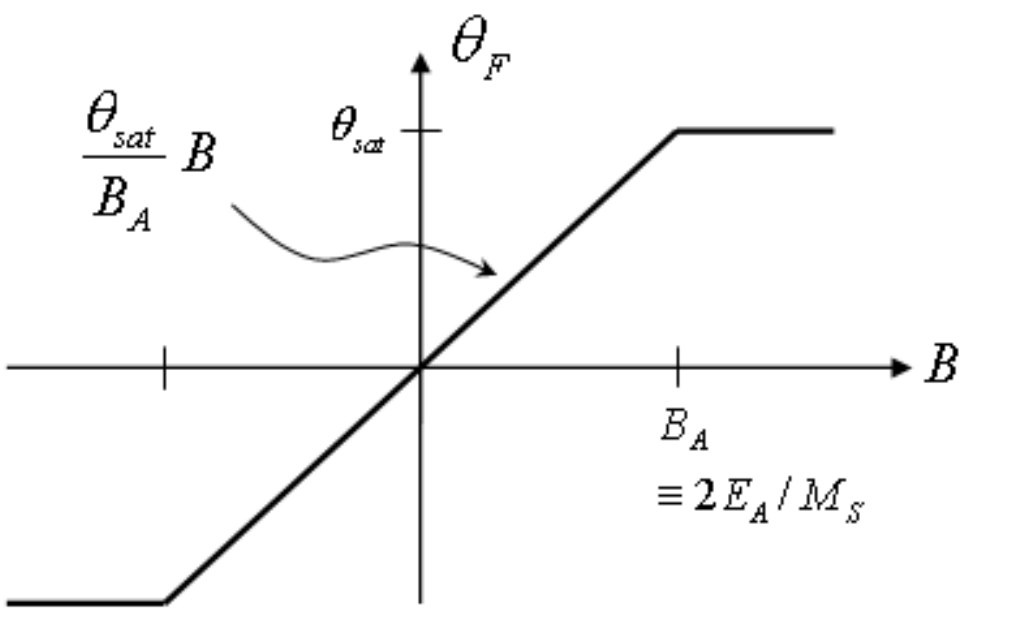}
   \caption{Faraday rotation in the indicator film for $B=B_{\perp}$ and $B_{\parallel}=0$.}
  \label{indicatorfig2}
 \end{center}
\end{figure}

In order to discuss the physics of Faraday-rotating indicator films, we first define a coordinate system where the indicator film is lying in the horizontal plane, see Fig. \ref{indicatorfig1} (a).
The interaction energy of the indicator film with the magnetic field $\textbf{B}$ is the sum of the anisotropy energy and the magnetostatic energy
\begin{equation}
 E_{int}=E_{A}{\sin}^2\phi-{{\textbf{B}\cdot\textbf{M}}_{s}}
 \label{filmenergy}
\end{equation}
which can be expanded as
\begin{equation}
 E_{int}=E_{A}{\sin}^2\phi- B_{\parallel} {M}_{s}\cos\phi- B_{\perp} {M}_{s}\sin\phi
 \label{filmenergy2}
\end{equation}
where $M_{s}$ is the spontaneous magnetization vector, related to the anisotropy energy and spontaneous in-plane field as $2E_{A} \equiv B_{A}M_{s}$. As seen in Fig. \ref{indicatorfig1} (b), there is symmetry around the vertical axis and configurations where the angle $\phi$ between the magnetization vector and the horizontal plane are identical and have the same energy. In order to find the equilibrium angle of the spontaneous magnetization, the expression is differentiated with respect to $\phi$ and set equal to 0. The case $B_{\parallel}=0$, $B_{\perp}=B$ is simple to treat analytically:
\begin{equation}
 \label{filmenergydiff1}
 0=(2E_{A}{\sin}\phi-B{M}_{s})\cos\phi
\end{equation}
Eq. \ref{filmenergydiff1} has the nontrivial solution
\begin{equation}
 \label{filmenergydiff2}
 \sin\phi=\frac{B{M}_{s}}{2E_{A}}=\frac{B}{B_{A}}
\end{equation}
The Faraday rotation is generally given by
\begin{equation}
 \theta_{F}=VtM_{s}\sin\phi=\theta_{sat}\frac{B}{B_{A}}
 \label{faradayrot}
\end{equation}
where $V$ is the Verdet constant of the film, $t$ is the thickness and the product $VtM_{s}\equiv \theta_{sat}$. The rotation is clearly proportional to the applied field inside a range defined by $\pm B_{A}$, outside of which the rotation saturates to $\theta_{sat}$. The rotation as a function of the magnetic field is plotted in Fig. \ref{indicatorfig2}.

\noindent \textit{Malus' law} is a relation from optics which gives the transmitted intensity $I$ of polarized light encountering a polarizer oriented along an angle relative to the polarization plane of the incoming light. For the case of a magneto-optical microscope with a relative polarizer-analyzer angle of $\theta = 90^{\circ}$ and a leakage intensity $I_{leak}$ through the polarizers, Malus' law reads
\begin{equation}
 I = I_{0}\cos^{2}(\theta + \theta_{F})+I_{leak} = I_{0}\sin^{2}(\theta_{F})+I_{leak}
 \label{malus}
\end{equation}
As the derivative of the squared sine function is at a minimum at an argument of $0^{\circ}$ and at a maximum at $45^{\circ}$, the configuration with a crossed polarizer and analyzer has the lowest sensitivity. Combining \eqs{faradayrot}{malus} gives the light intensity as a function of applied field $B_a$
\begin{equation}
 I = I_{0}\sin^{2}(\theta_{sat}\frac{B}{B_{A}})+I_{leak}
 \label{intensity1}
\end{equation}
which can be solved for
\begin{equation}
 B = \frac{B_{A}}{\theta_{sat}}\arcsin{\sqrt{\frac{I - I_{leak}}{I_{0}}}}
 \label{intensity2}
\end{equation}
This approximation is applicable for $B_{\parallel} \ll B_{A}$. A more general expression for the light intensity in the case of general $\theta$ is
\begin{equation}
 I = I_{0}\sin^{2}(\theta_{sat}\frac{B}{B_{A}}+\theta)+I_{leak}
 \label{intensity3}
\end{equation}
giving a corresponding equation for the magnetic field
\begin{equation}
 B = \frac{B_{A}}{\theta_{sat}}(\arcsin{\sqrt{\frac{I - I_{leak}}{I_{0}}} - \theta})
 \label{intensity4}
\end{equation}
The most general case with a significant $B_{\parallel}$ is in theory also analytically solvable, but requires the solution of a quartic equation. With an EuSe indicator film, one would get a similar functional dependency, but in this case the Faraday rotation angle is just the product of the Verdet constant and the thickness, so the coefficient in front of the parenthesis in Eq. \ref{intensity4} would instead be $1/Vt$.

\section{Calibration} \label{calibrationsection}

\begin{figure}[h!]
 \begin{center}
  \includegraphics[width=0.6 \linewidth]{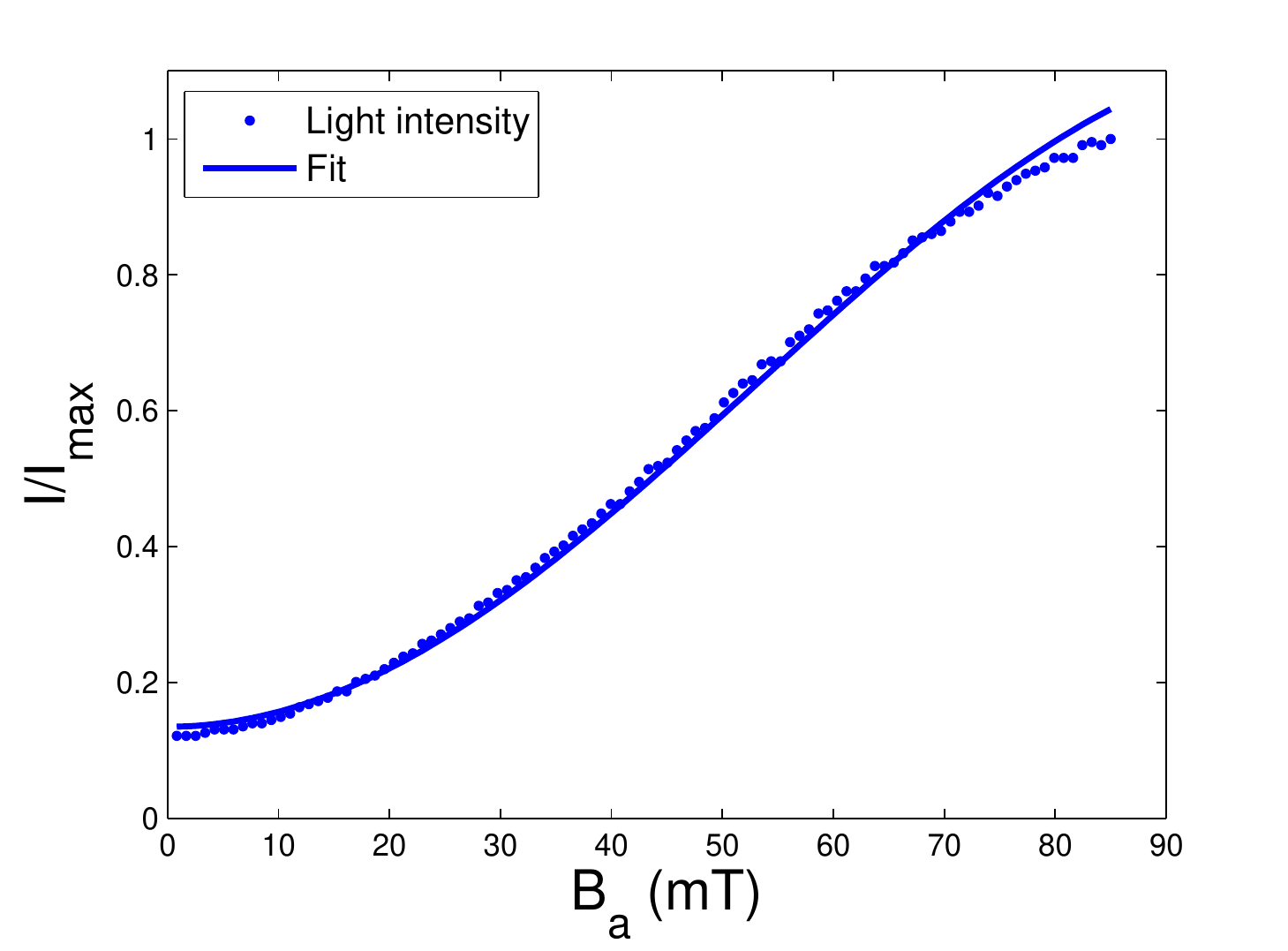}
   \caption{Typical relationship between intensity and magnetic field for a pixel. A fit to Eq. \ref{intensity1} is shown with $I_{leak}/I_{0}$ = 0.135 and $\theta_{sat}/B_{A}$ = 14.9 $T^{-1}$. $\theta$ was fixed to zero, corresponding to the used crossed polarizer-analyzer configuration.}
  \label{calib}
 \end{center}
\end{figure}
\noindent As a light source often does not have a uniform beam and the intensity of the Faraday rotated light is not a linear function of the local magnetic field, the raw images must be calibrated if the local values of the magnetic fields and current densities are to be deduced. The light intensity for crossed polarizers is an sigmoid function of $B_{a}$, as seen in Fig. \ref{calib}. In general, the best possible fit function depends on the ratio of the saturation field $B_A$ of the indicator relative to the maximum applied field $B_a$. The large temperature range of the ferrite garnets allows the light response of the indicator film as a function of magnetic field to be recorded above the $T_{c}$ of the superconductor. A series of images is captured with increasing applied magnetic field $B_a$ and the light intensity values $I(x,y)$ are fit to deduce the function $I(B)$ for every pixel. This function is given in Eq. \ref{intensity3} for the case of a large range in $B_a$, for small $B_a$ one would instead use the quadratic approximation $I = aB_a^2 + bB_a + c$. Raw images of the light intensity $I(x,y)$ captured below $T_{c}$ can then be calibrated into images of $B_{z}(x,y)$ from the known, monotonic relation $B(I)$, which would be either Eq. \ref{intensity4} for large $B_a$ or the solution of the stated quadratic equation for $B$ in the case of small $B_a$.

It is here assumed that the light response of the indicator film is approximately independent of the temperature at temperatures at and below the $T_{c}$ of the material in question. After calibrating a magneto-optical image or image series, one should check that the magnetic field values far outside the sample converge to $B_{a}$. An example of such a consistency check is shown in Fig. \ref{bjprof} (a) in chapter \ref{magnetic}.

MATLAB codes for calibration with a sigmoid function and a quadratic function are included in {\color{red}sigmacal.m} and {\color{red}powercal.m}, respectively.

\chapter{Inversion of the Biot-Savart law} \label{biotsavartchapter}
A calibrated magneto-optical image can be used to deduce the corresponding current distribution by inversion of the Biot-Savart law, as it is a local relation in Fourier space. In order to proceed with this problem, the mathematics of the Biot-Savart law and Fourier transforms will first have to be discussed.
\section{Biot-Savart law in Fourier space}
\begin{figure}[h!]
 \begin{center}
   \includegraphics[width=0.57 \linewidth]{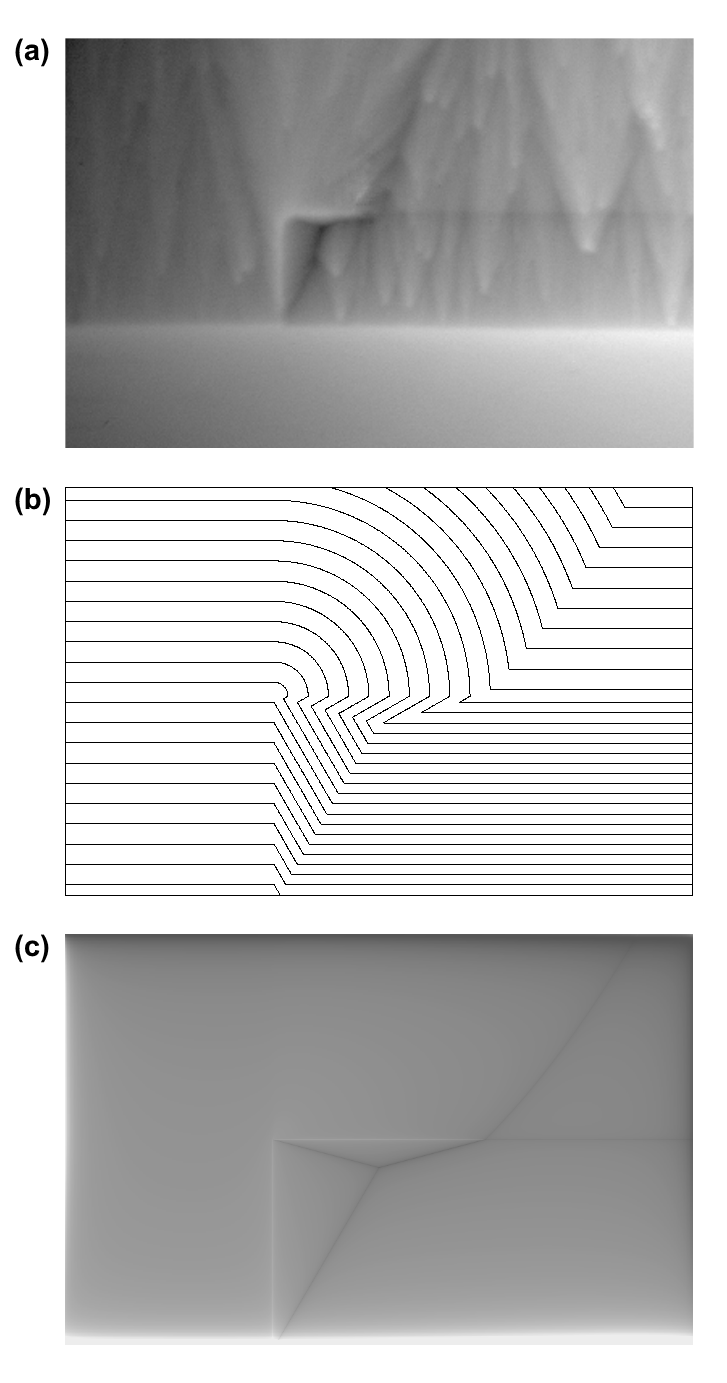}
   \caption{(a) Magneto-optical image of a MgB$_{2}$ sample with gold coating captured at $T = 3.5$ K and $B_{a}$ = 65.1 mT. The gold coated area has a rectangular shape and is situated in the lower right half of the sample. (b) Current lines compatible with the magneto-optical image in (a). (c) Calculated magneto-optical image from the currents in (b).}
  \label{mgb2fig}
 \end{center}
\end{figure}

The Biot-Savart law reads in full generality
\begin{equation}
\textbf{B(r)} = \frac{\mu_{0}}{4\pi} \int \frac{\textbf{j}(\textbf{r})' \times (\textbf{r} - \textbf{r}')}{|\textbf{r} - \textbf{r}'|^3} d^{3} \textbf{r}'
\label{biotr}
\end{equation}
The three components of the magnetic flux density generated by a three-dimensional current distribution can thus be written
\begin{equation}
B_{x}(x,y,z) = \frac{\mu_{0}}{4\pi} \int \int \int \frac{j_{y}(x',y',z')(z - z') - j_{z}(x',y',z')(y - y')}{\sqrt{[ (x - x')^2 +(y - y')^2 +(z - z')^2 ]^3}} dx'dy'dz'
\label{biotx}
\end{equation}

\begin{equation}
B_{y}(x,y,z) = \frac{\mu_{0}}{4\pi} \int \int \int \frac {j_{z}(x',y',z')(x - x') - j_{x}(x',y',z')(z - z')}{\sqrt{[ (x - x')^2 +(y - y')^2 +(z - z')^2 ]^3}} dx'dy'dz'
\label{bioty}
\end{equation}

\begin{equation}
B_{z}(x,y,z) = \frac{\mu_{0}}{4\pi} \int \int \int \frac {j_{x}(x',y',z')(y - y') - j_{y}(x',y',z')(x - x')}{\sqrt{[ (x - x')^2 +(y - y')^2 +(z - z')^2 ]^3}} dx'dy'dz'
\label{biotz}
\end{equation}
The ``forward" calculation of the magnetic field from a three-dimensional current distribution is always possible. An example is shown below: In a series of MgB$_{2}$ samples with half of the rim covered with a gold layer, magneto-optical images suggest a higher critical current density $j_{c,1}$ in the gold-covered area as opposed to the $j_{c,2}$ in the sample area without gold, probably due to sample degradation in the uncovered area. A representative example of this behavior is shown in Fig. \ref{mgb2fig} (a). Current conservation limits possible flow patterns the one shown in Fig. \ref{mgb2fig} (b) and the resulting field distribution and magneto-optical image from a ratio $j_{c,1}$/$j_{c,2} = 2$ was calculated. It is seen that the calculated image in Fig. \ref{mgb2fig} (c) has the same d-lines as the observed image in Fig. \ref{mgb2fig} (a), showing that this current flow pattern is indeed qualitatively consistent with the observed image.

In contrast, the inverse problem does not in general have a unique solution. For samples with $d<2\lambda$, it is a good approximation to consider the in-plane current density to be constant in the $z$-direction \cite{Brandt2, Brandt3}. For thicker samples, the measured current densities must be regarded as averages over the thickness, $j(x,y) = \frac{1}{d}\int_{-d/2}^{d/2}j(x,y,z)dz$. Such an average is called a \textit{sheet current density}. It is also assumed in the following that the current has no significant $j_{z}$-component, again a good approximation for thin films. These assumptions result in the following equations for the magnetic field components in a height $h$ above a thin sample with thickness $d$
\begin{equation}
B_{x}(x,y,h) = \frac{\mu_{0}}{4\pi} \int_{-d/2}^{d/2} \int \int \frac{j_{y}(x',y') h }{\sqrt{[ (x - x')^2 +(y - y')^2 + h^2 ]^3}} dx'dy'dz'
\label{biotxh}
\end{equation}

\begin{equation}
B_{y}(x,y,h) = \frac{\mu_{0}}{4\pi} \int_{-d/2}^{d/2} \int \int \frac { - j_{x}(x',y')h}{\sqrt{[ (x - x')^2 +(y - y')^2 + h^2 ]^3}} dx'dy'dz'
\label{biotyh}
\end{equation}

\begin{equation}
B_{z}(x,y,h) = \frac{\mu_{0}}{4\pi} \int_{-d/2}^{d/2} \int \int \frac {j_{x}(x',y')(y - y') - j_{y}(x',y')(x - x')}{\sqrt{[ (x - x')^2 +(y - y')^2 + h^2 ]^3}} dx'dy'dz'
\label{biotzh}
\end{equation}
As shown by Roth \cite{Roth}, it is possible to calculate a two-dimensional current distribution from the a two-dimensional image of $B_{z}$ measured in a fixed distance from the film. Roth utilized the fact that the Biot-Savart law has translational symmetry and that the convolution theorem therefore can be applied. The Green's function
\begin{equation}
G(x - x',y - y',h) = \frac{\mu_{0}}{4\pi} \frac {h}{\sqrt{[ (x - x')^2 +(y - y')^2 + h^2 ]^3}}
\label{greens}
\end{equation}
has an analytical Fourier transform
\begin{equation}
\tilde{G}(k_{x},k_{y},h) = \frac{\mu_{0}}{2}e^{- h \sqrt{k^{2}_{x} + k^{2}_{y}}}
\label{greensf}
\end{equation}
Using this transform, the two-dimensional Fourier transforms of Eq. \ref{biotxh}, Eq. \ref{biotyh} and Eq. \ref{biotzh} can be written

\begin{equation}
\tilde{B}_{x}(k_{x},k_{y},h) = \frac{\mu_{0} d}{2} \tilde{j}_{y}(k_{x},k_{y})e^{- hk}
\label{biotxhf}
\end{equation}

\begin{equation}
\tilde{B}_{y}(k_{x},k_{y},h) = -\frac{\mu_{0} d}{2} \tilde{j}_{x}(k_{x},k_{y})e^{- hk}
\label{biotyhf}
\end{equation}

\begin{equation}
\tilde{B}_{z}(k_{x},k_{y},h) = i\frac{\mu_{0} d}{2} (\frac{k_{y}}{k} \tilde{j}_{x}(k_{x},k_{y}) - \frac{k_{x}}{k} \tilde{j}_{y}(k_{x},k_{y}))e^{- hk}
\label{biotzhf}
\end{equation}
where $k = \sqrt{k^{2}_{x} + k^{2}_{y}}$ and the integration over $dz'$ has been carried out through a simple multiplication with the film thickness $d$. The equation Eq. \ref{biotzhf} is simplified when considering the magnetization scalar field $g(\textbf{r})$, which relates to the sheet current as
\begin{equation}
 j(x,y) = \nabla \times \hat{z} g
 \label{mag}
\end{equation}
As the divergence of the curl of a scalar field always is zero, this form ensures that the current density has no divergence, which again is required by current conservation inside a superconductor with no external current sources
\begin{equation}
 \nabla \cdot j(x,y) = \nabla \cdot (\nabla \times \hat{z} g) = 0
 \label{mag2}
\end{equation}
In components Eq. \ref{mag} can be written $j_{x}(x,y) = \partial g(x,y)/ \partial y $ and $j_{y}(x,y) = - \partial g(x,y)/ \partial x$. In Fourier space, these partial derivatives are performed by multiplying $\tilde{g}(\textbf{k})$ by $ - ik_{y}$ and $ - ik_{x}$, respectively. This enables us to rewrite Eq. \ref{biotzhf} as
\begin{equation}
\tilde{B}_{z}(k_{x},k_{y},h) = \frac{\mu_{0} d}{2} (\frac{k^2_{y}}{k} \tilde{g}(\textbf{k}) + \frac{k^2_{x}}{k} \tilde{g}(\textbf{k}))e^{- hk}
\label{biotzhfs}
\end{equation}
which can be simplified to
\begin{equation}
 \tilde{B}_{z}(k_{x},k_{y},h)=\frac{\mu_{0}kd}{2}\tilde{g}(\textbf{k})e^{-hk}
 \label{fbs}
\end{equation}
Thus, the magnetic field can then be calculated from the magnetization, and this is especially simple in Fourier space, where one can see that the Biot-Savart law for $B_{z}$ is a local relation between $B_z$ and $g$.

For the case $d << \lambda$, Jooss \cite{Jooss2} treated the finite thickness of the sample more accurately by integrating over the $dz'$ instead of just multiplying with the sample thickness $d$, which results in $kd/2$ being replaced with $sinh(kd/2)$ in Eq. \ref{fbs}. This can be seen from keeping the $z-z'$ term in Eq. \ref{biotz} so that $h$ will be shifted to $h-z'$ in Eqs. \ref{biotzh}, \ref{greens} and \ref{greensf}, noting that the current density terms have no $z$-dependence after averaging over this dirction and observing that the last integration in Eq. \ref{biotzh} over $dz'$ will pick up the factor $(e^{kd/2} - e^{-kd/2})$ which together with the prefactor $1/2$ in Eq. \ref{greensf} gives $sinh(kd/2)$. Again, for a thin sample, $sinh(kd/2)$ reduces to the previous result $kd/2$ by series expansion.

For the Bean model, the magnetization functions are straight planes with slope $j_{c}$. One should also note the product of the height above the sample and the wave number in the exponential in Eq. \ref{fbs}. This tells us that $B_{z}$ is a low-pass filtered $g$. Proximity of the indicator film to the current sheet is thus important to avoid blurring of the image of $B_{z}$. If one wants to go the other way, i.e. obtain the currents from a measurement of $B_{z}$, one solves Eq. \ref{fbs} for $\tilde{g}(\textbf{k})$
\begin{equation}
 \tilde{g}(\textbf{k})=\frac{2}{\mu_{0} kd}\tilde{B}_{z}e^{hk}
 \label{fbsi}
\end{equation}
and then the currents are found from $j_{x}(x,y) = \partial g(x,y)/ \partial y $ and $j_{y}(x,y) = - \partial g(x,y)/ \partial x$. The absolute value of the current density is then found by $|j| = \sqrt{j^2_{x}+j^2_{y}}$. This inversion can be done for experimentally obtained magneto-optical images after they have been calibrated into two-dimensional maps of $B_{z}(x,y)$.

\section{Fast Fourier transformation}
In order to implement these calculations numerically, a quick detour to Fourier transformations is needed. The Fourier transformation and its inverse are defined by

\begin{equation}
\tilde{f}(k) = \int^{\infty}_{-\infty}  dx f(x) e^{-ikx}
\label{ftk}
\end{equation}

\begin{equation}
{f}(x) = \frac{1}{2\pi} \int^{\infty}_{-\infty}  dk \tilde{f}(k)e^{ikx}
\label{ftx}
\end{equation}
\textit{The Discrete Fourier Transform} (DFT) is the numerical discretization of the Fourier transform and its inverse. On a grid of size $L$ with $N$ points, they are defined by

\begin{equation}
\tilde{g}_{n} = \sum\limits_{j=0}^{N-1}  g_{j} e^{-i \frac{2 \pi jn}{N}}
\label{gn}
\end{equation}

\begin{equation}
g_{j} = \frac{1}{N} \sum\limits_{n=0}^{N-1} \tilde{g}_{n} e^{i \frac{2 \pi jn}{N}}
\label{gj}
\end{equation}
where the continuous integration variables are replaced with discrete counterparts: $x \rightarrow x_{n} = Ln/N$ and $k \rightarrow k_{j} = 2 \pi j/L$. As a consequence of the Nyquist-Shannon theorem, the Brillouin zones will have to be rearranged to yield correct results, so in practice
\begin{equation}
k_{j} = \frac{2\pi}{L}(j-\frac{N}{2})
\label{brillouin}
\end{equation}
The convolution theorem states that the Fourier transform of a convolution of functions is the product of Fourier transforms point by point
\begin{equation}
\int^{\infty}_{-\infty} dx' f(x') g(x - x') = \frac{1}{2\pi} \int^{\infty}_{-\infty}  dk \tilde{f}(k)\tilde{g}(k)e^{ikx}
\label{convolution1}
\end{equation}
which in a discrete form reads
\begin{equation}
\sum\limits_{l=0}^{N-1}  g_{l} h_{j - l}  = \frac{1}{N} \sum\limits_{n=0}^{N-1} \tilde{g}_{n} \tilde{h}_{n} e^{i \frac{2 \pi jn}{N}}
\label{convolution2}
\end{equation}
A DFT requires $\mathcal{O}(N^2)$ operations for each dimension. \textit{The Fast Fourier Transform} (FFT) is any fast algorithm to compute the DFT of a function. The two-dimensional FFT requires $N_{x}N_{y}(1 + 2log(N_{x}N_{y}))$ operations where $N_{x}$ and $N_{y}$ are the dimensions of the grid. This is obviously a huge improvement over the DFT, which would require $\mathcal{O}((N_{x}N_{y})^2)$ operations for a two dimensional grid.

A DFT is periodic, and if one tries to sample a function that is not periodic, this will introduce an error. As a consequence, one should ensure periodic boundary conditions when sampling magneto-optical images by for example including the indicator film around it. The finite size of an image also introduces artifacts in the form of a superlattice of spurious current distributions. The imaged area should be at least a factor of two larger than the sample to reduce this effect.

It is also possible to invert the Biot-Savart law with matrix inversion. This approach requires $\mathcal{O}((N_{x}N_{y})^3)$ operations. The Toeplitz symmetry of the integral kernel of the Biot-Savart law has been utilized by Wijngaarden et al. to reduce the number of operations to $\mathcal{O}((N_{x}N_{y})^{2.25})$ \cite{Wijngaarden}.

\section{Magnetic field to current inversion} \label{magnetic}

\begin{figure}[h!]
 \begin{center}
  \includegraphics[width=1 \linewidth]{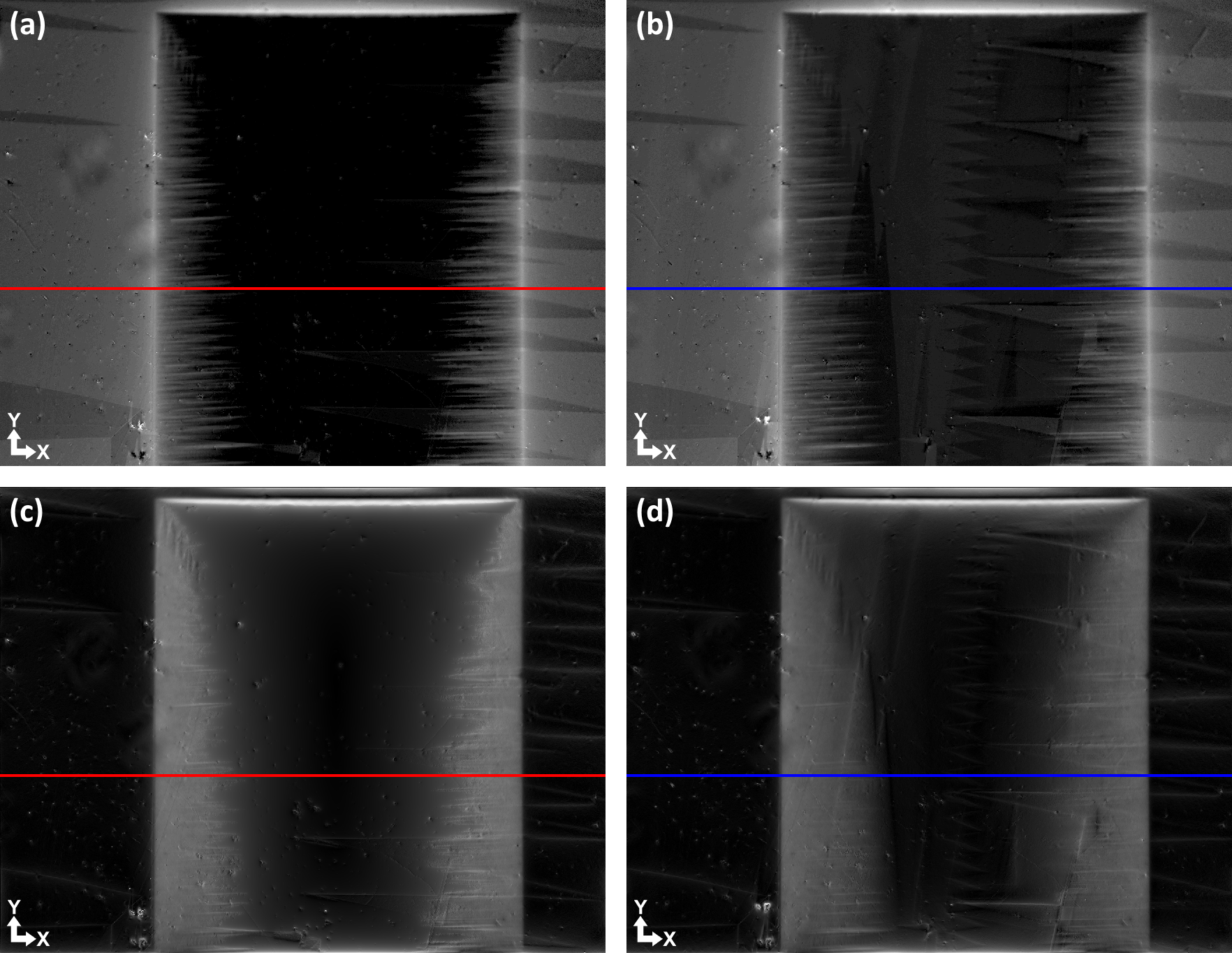}
   \caption{Magnetic field maps and current density maps. Magnetic field map (a) is not significantly perturbed by
   domains, while map (b) is. The corresponding current density maps are (c) and (d). The measurements are obtained with $B_{a}$ = 8.5 mT
   and $T$ = 4 K on a YBCO film on a $14^{\circ}$ tilted substrate. The superconducting strip is approximately 0.9 mm wide.}
  \label{domains}
 \end{center}
\end{figure}

\noindent Having discussed the physics of in-plane magnetized ferrite garnet indicator films, calibration and the mathematics of FFTs, it is now time to proceed to numerical inversion of the Biot-Savart law. In theory, it should be simple to extract current density distributions from the calibrated magneto-optical images by applying the inverse Biot-Savart law in Fourier space and calculating the magnetization $g$ and differentiating it to get $j_x$ and $j_y$.

\begin{figure}[t!]
 \begin{center}
  \includegraphics[width=0.61 \linewidth]{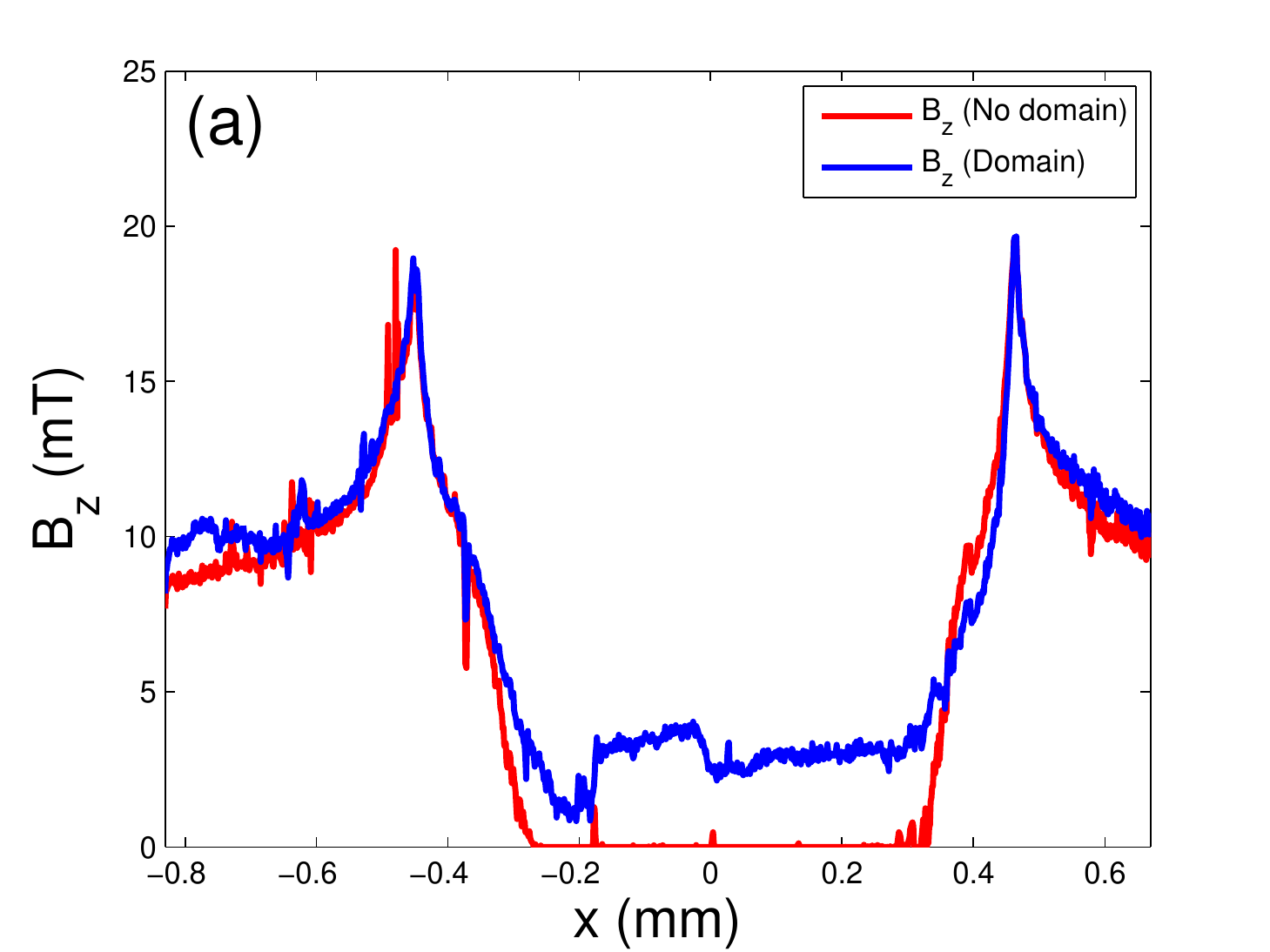}
  \includegraphics[width=0.61 \linewidth]{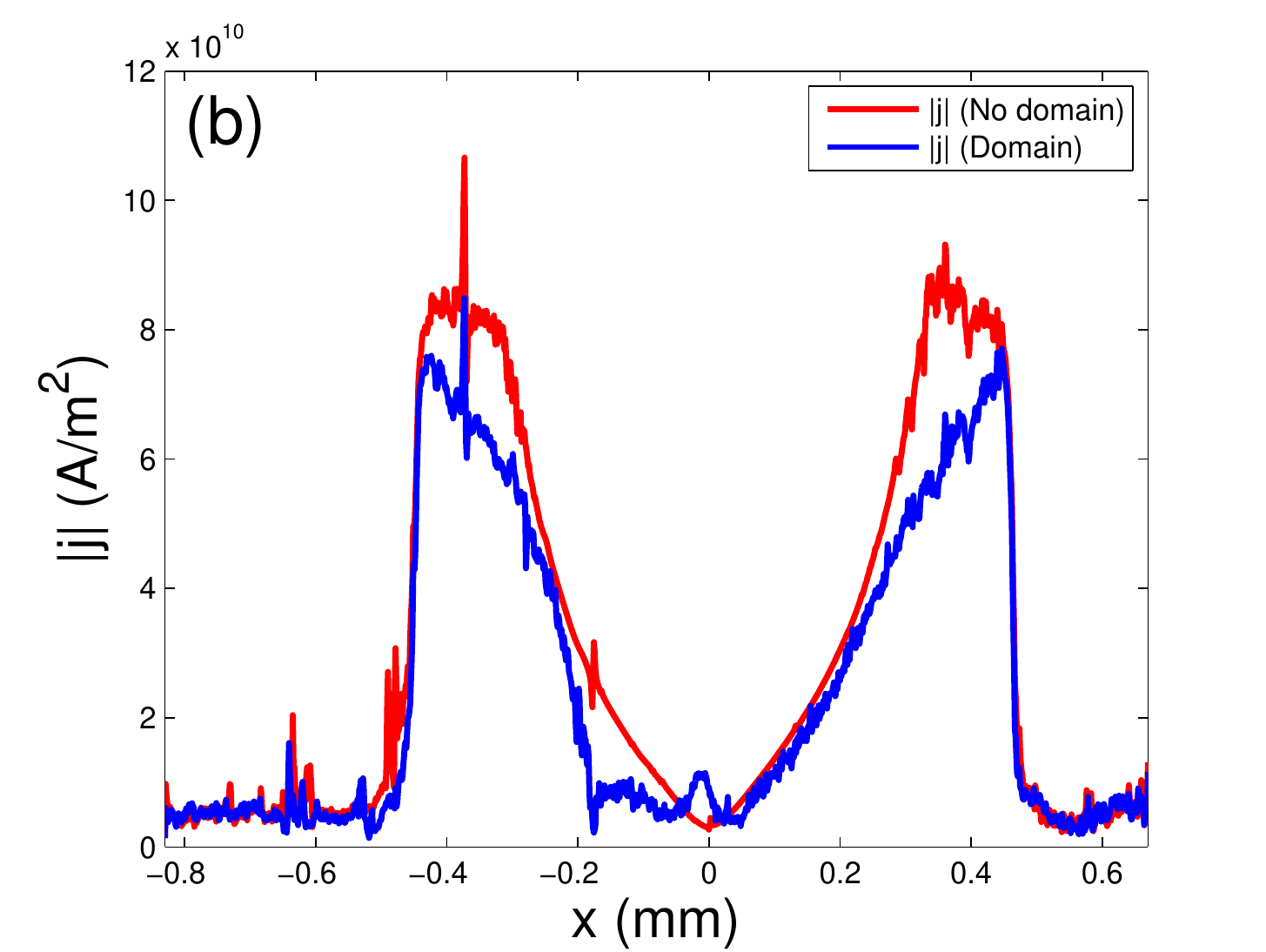}
  \caption{(a) Cross-sections of magnetic field maps (a) and (b) in Fig. \ref{domains}. (b) Cross-sections of current density maps (c) and (d) in Fig. \ref{domains}. The measurements are obtained with $B_{a}$ = 8.5 mT and $T$ = 4 K on a YBCO film on a $14^{\circ}$ tilted substrate.}
  \label{bjprof}
 \end{center}
\end{figure}

In practice, there are several problems. Defects in the superconducting film and in the magneto-optical indicator perturb the magnetic field and the current pattern is sensitive to even small perturbations of this kind. Even worse are sawtooth-shaped magnetic domain boundaries in the indicator film, but they can be removed, or at least moved by application of a small in-plane field. As mentioned, the indicator film is also sensitive to the in-plane components $B_{x}$ and $B_{y}$, which non-locally reduces Faraday rotation and causes an underestimation of $B_{z}$. This problem is most pronounced for strong induced currents resulting from a high $B_a$. It manifests itself as an unphysical current outside the sample and sharp, unphysical peaks in the current density close to the sample edge. At low $B_a$, a ``na{\"i}ve" inversion, disregarding the in-plane component, will give a good approximation to the actual current density distribution. In Figs. \ref{domains} (a) and (b), maps of magnetic field values $B_{z}$ obtained by calibration of raw images captured at $B_a$ = 8.5 mT and $T$ = 4 K of a $d = 200$ nm thick YBCO film grown on a $14^{\circ}$ tilted substrate are shown. The corresponding inverted current densities $|j|$ are shown in Figs. \ref{domains} (c) and (d), respectively.

Cross sections of the $B_{z}$ maps and the $|j|$ maps in Fig. \ref{domains} are shown in Figs. \ref{bjprof} (a) and \ref{bjprof} (b), respectively. Both unperturbed cross sections of Figs. \ref{domains} (a) and (c) and cross sections of Figs. \ref{domains} (b) and (d) perturbed by a domain are shown. The images were captured at a relatively low $B_a$, so the in-plane corrections are small and the na{\"i}ve inversion scheme reproduces a Bean model-like current profile from a Bean model-like magnetic field profile, compare with Figs. \ref{thinfilm} (a) and (b). Note that $|j(x)|$, not $|j_y(x)|$ is plotted in Fig. \ref{bjprof}, however these coincide almost perfectly far away from the end of the strip. It is also seen in Figs. \ref{bjprof} (a) and \ref{bjprof} (b) that the $B_{z}$ maps and $|j|$ maps are severely perturbed by domain patterns in the indicator film. Experiments have shown that it is possible to use the na{\"i}ve inversion at higher magnetic fields by use of an iterative algorithm \cite{Laviano}.

From Eq. \ref{c9}, one can calculate $j_c$ from the applied field $B_a$, sample thickness $d$, sample width $2w$ and penetration depth $L$ for the case of the image in Fig. \ref{domains} (a), where the sample width $2w$ = 394 pixels and a reasonable value for the depth of the flux front away from the corners is $L = 100$ pixels. One starts with noting that the flux penetration depth $L$ on both sides of the flux-free region of width $2a$ must sum up to $2w$, or $w = L + a$ by symmetry. Combining this expression with Eq. \ref{c9} gives

\begin{equation}
w = L + \frac{w}{cosh(\frac{B_{a} \pi }{\mu_{0} j_{c} d})}
 \label{jestimate1}
\end{equation}

\noindent Eq. \ref{jestimate1} is then rearranged as to isolate $j_c$

\begin{equation}
 j_{c} = \frac{B_{a} \pi}{acosh(\frac{w}{w-L})\mu_{0}d}
 \label{jestimate2}
\end{equation}
Inserting $B_{a}$ = 8.5 mT, $\mu_{0} = 4\pi \cdot 10^{-7}$ N/A$^2$, $d = 200$ nm, $w = 197$ pixels and $L = 100$ pixels in Eq. \ref{jestimate2}, we obtain $j_{c} = 7.96 \cdot 10^{10}$ A/m$^2$, in good agreement with the flat regions in the unperturbed current density profile in Fig. \ref{bjprof} (b). One can also see from Fig. \ref{bjprof} (a) that the calibration is reasonable, as the $B_z$-values fall off towards $B_a = 8.5$ mT outside the sample. MATLAB code for inversion is included in {\color{red}inversion.m}.

\section{Visualizing current lines} \label{currentlines}

\begin{figure}[h!]
 \begin{center}
  \includegraphics[width=0.495 \linewidth]{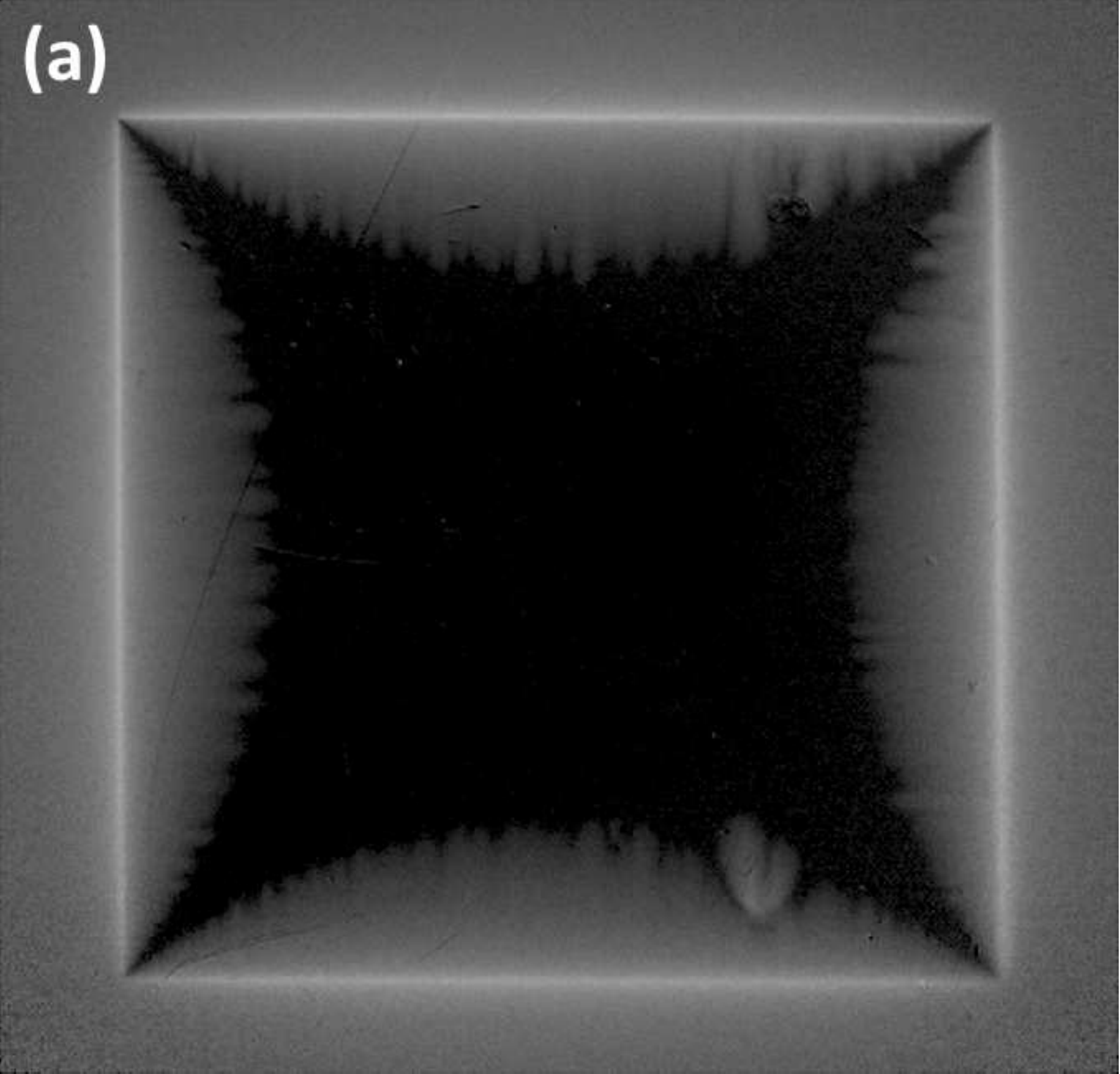}
  \includegraphics[width=0.495 \linewidth]{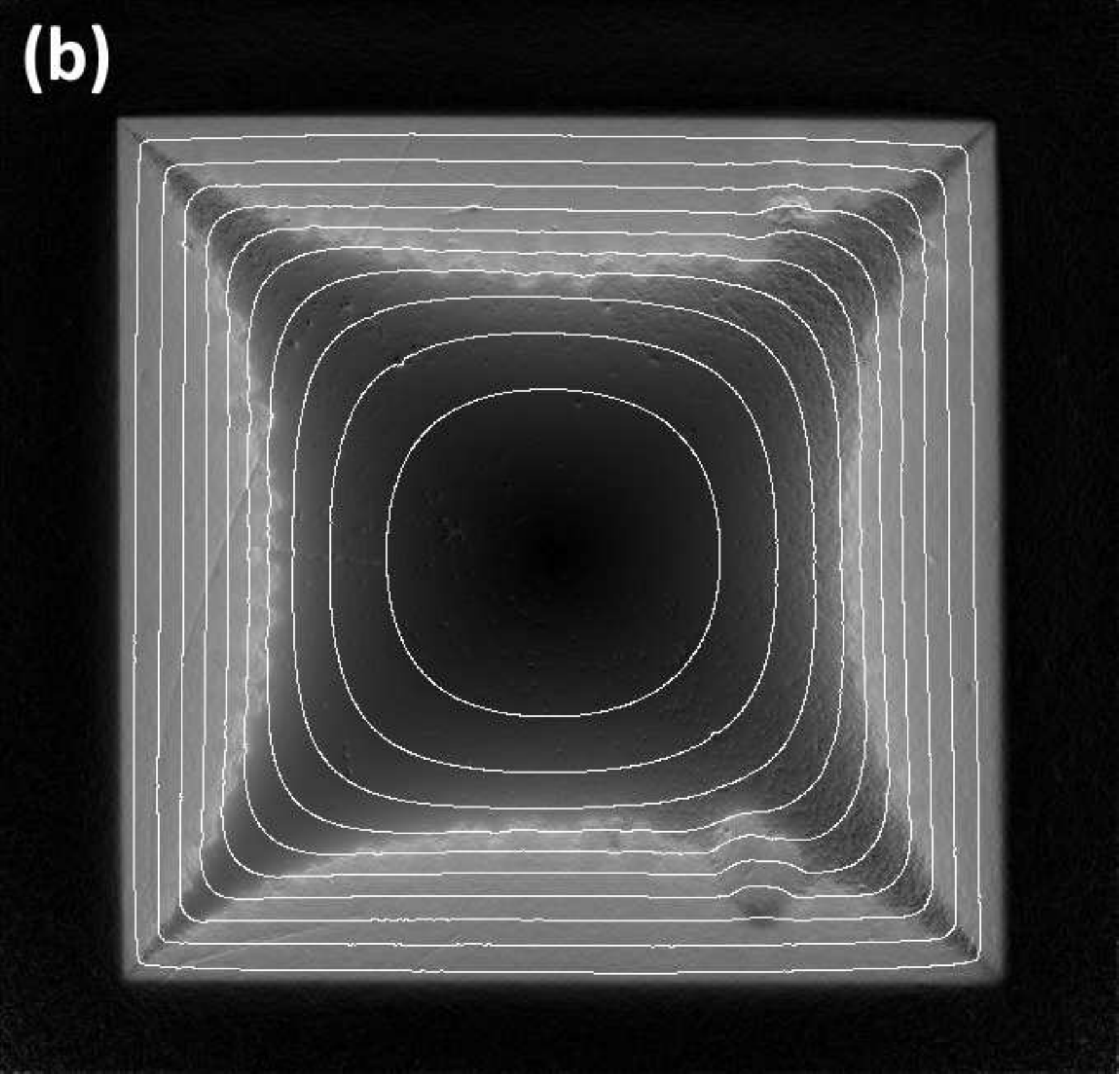}
  \caption{Calibrated magneto-optical image of a 420 nm thick NbN-AlN-NbN trilayer of dimension 4.13x4.02 mm$^{2}$ (a) and corresponding current density with superimposed current lines (b). Image (a) was captured at $B_{a}$ = 13.2 mT and $T$ = 3.5 K after zero-field cooling.}
  \label{bzandcurrent}
 \end{center}
\end{figure}

\noindent The visualization of current distributions resulting from inversion of magneto-optical images can be improved by adding current lines. This can be done by applying the \textit{contour} function in MATLAB to the magnetization $g$. To see why this works, consider the following argument: The gradient of $g$ is normal to the contour lines of $g$, that is, lines where $g$ is constant. Dot the current vector $\textbf{j} = j_x \hat{x} + j_y \hat{y}$ with $\nabla g = (dg/dx) \hat{x} + (dg/dy) \hat{y}$ to get $\textbf{j} \cdot \nabla g = 0$ as $j_{x}(x,y) = \partial g(x,y)/ \partial y $ and $j_{y}(x,y) = - \partial g(x,y)/ \partial x$. The current thus has no component transverse to the contour lines of $g$ and any current must therefore flow along them.

As an illustration, shown in Fig. \ref{bzandcurrent} (a) is a calibrated magneto-optical image of a 420 nm thick NbN-AlN-NbN trilayer captured at $B_{a}$ = 13.2 mT and $T$ = 3.5 K, while the corresponding current density map with superimposed current lines is shown in Fig. \ref{bzandcurrent} (b) (the non-superconducting AlN interlayer is very thin). MATLAB code for plotting current lines is included in {\color{red}currentlines.m}.

\newpage
\section{Magnetometry with Faraday magneto-optical imaging} \label{Magnetometry}

The magnetic moment of the magnetization of a sample with volume $V$ is defined as the following integral

\begin{equation}
m = \int_V g dxdydz
\label{moment}
\end{equation}

\noindent As the magnetization $g$ is derived from ${B}_{z}$ by a Fourier transformation and Eq. \ref{fbsi}, it should be of little surprise that it is possible to do magnetometry with Faraday magneto-optical imaging \cite{Qureishy}. A practical implementation of this procedure requires properly calibrated magneto-optical images that must again cover a significantly larger area than the sample in order to yield correct results. The induced fields on the edges of a thin-film superconductor may be significantly higher than $B_a$, as depicted in Fig. \ref{thinfilm} (b). In this case, the field range of the calibration images should be at least twice as large as the field range of $B_a$ in order to not introduce clipping of the highest induced field values. Hysteresis loops can be calibrated by using a sigmoid function centered at $B_a$ = 0, the calibration function thus having monotonic behavior in both the positive and negative $B_a$ ranges. Calibrated images should in this case be saved in the MATLAB {\color{red}.mat} format as the commonly used {\color{red}.tif} format for camera output cannot hold negative values.

Also required for the magnetometry procedure is an ordinary optical image of the sample with intensity values binarized into zeros outside the sample and ones inside the sample. This image is used to remove the field contribution outside the sample and corresponds to the $x$ and $y$ limits of the integral in Eq. \ref{moment}. In the attached MATLAB code, the $z$-integration is performed by not dividing by $d$ in \ref{fbsi} in the first place.

\begin{figure}[h]
 \begin{center}
  \includegraphics[width=0.6 \linewidth]{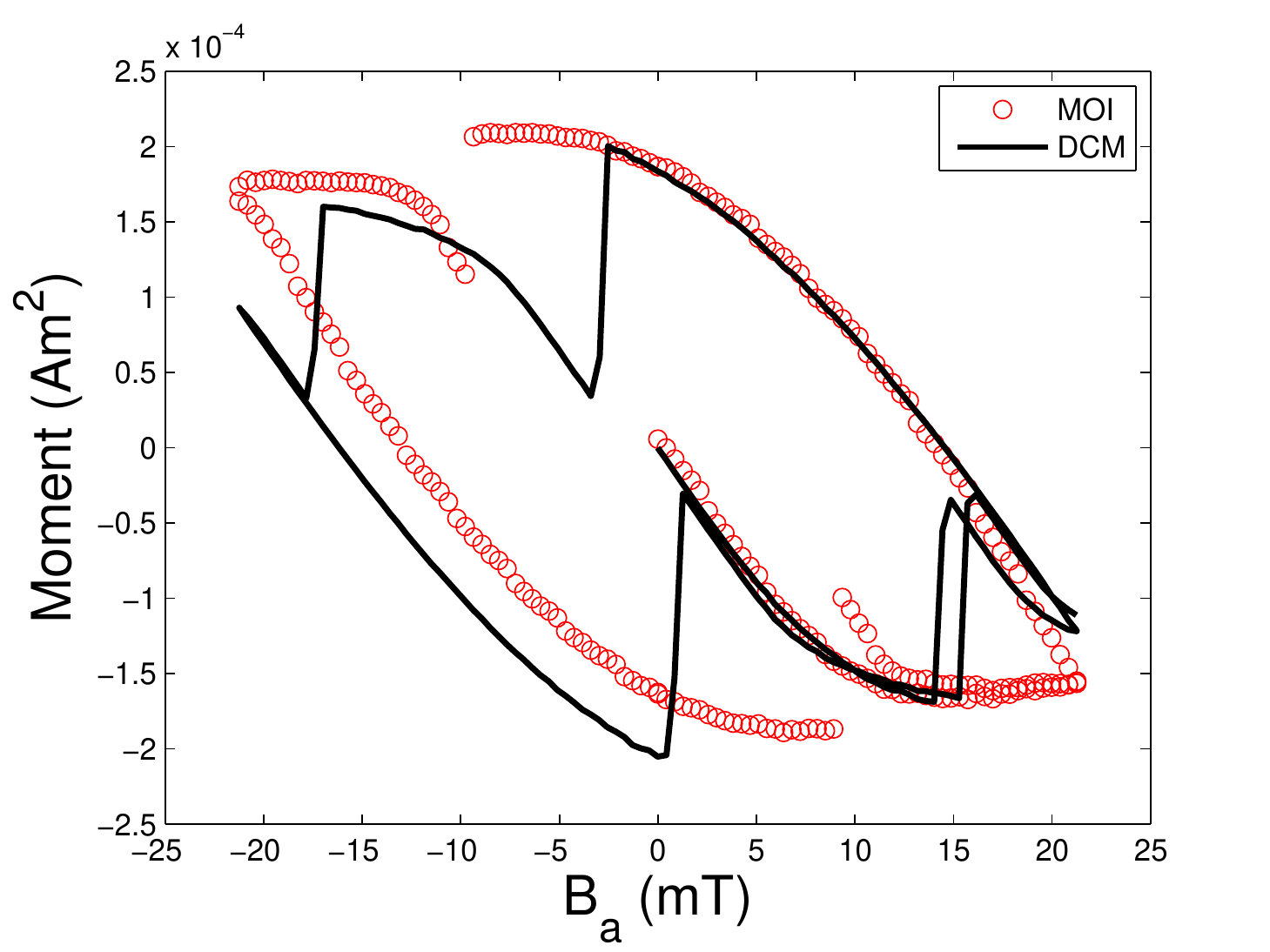}
  \caption{Magnetic hysteresis loop of an NbN-AlN-NbN trilayer from magneto-optical imaging, together with DCM data serving as a consistency check. Both measurements were done at $T$ = 5 K for a zero-field cooled sample. Several flux jumps are visible as abrupt discontinuities in the curves.}
  \label{momentloop}
 \end{center}
\end{figure}

\noindent An example of such a calculation for the NbN-AlN-NbN trilayer shown in Fig. \ref{bzandcurrent} (a) and (b) is shown in Fig. \ref{momentloop}. DC magnetometry (DCM) data is included, confirming the validity of the corresponding magneto-optical procedure. Abrupt discontinuities in the curves correspond to flux jumps. The flux jumps are of a stochastic nature and thus the flux jumps do not overlap exactly in $B_a$, and also cause the hysteresis curves to be partially shifted. However, the virgin curves agree very well before the DCM curve has a flux jump, and the agreement is also good in the second branch of the hysteresis curve. MATLAB code for calculating magnetic moments and plotting a hysteresis loop is included in {\color{red}momentloop.m}.

%\clearpage

\end{document}